\documentclass[english,
]{aa}
\usepackage{epsf,amsfonts,amssymb,graphicx,fancyheadings,caption,float}
\usepackage{natbib}
\usepackage{babel}
\usepackage{txfonts}
\usepackage{rotatingori}

\bibpunct{(}{)}{;}{a}{}{,}   

\begin{document}

\title{Tracking down R Coronae Borealis stars from their mid-infrared WISE colours
}

\author{
P.~Tisserand
}

\institute{
Research School of Astronomy and Astrophysics, Australian National University, Cotter Rd, Weston Creek
 ACT 2611, Australia
}

\offprints{Patrick Tisserand; \email{tisserand@mso.anu.edu.au}}

\date{}

\abstract {R Coronae Borealis stars (RCBs) are hydrogen-deficient and carbon-rich supergiant stars. They are very rare, as only $\sim50$ are actually known in our Galaxy. Interestingly, RCBs are strongly suspected to be the evolved merger product of two white dwarfs and could therefore be an important tool to understand Supernovae type Ia in the double degenerate scenario. Constraints on the spatial distribution and the formation rate of such stars are needed to picture their origin and test it in the context of actual population synthesis results.}
{It is crucial to increase significantly the number of known RCBs. With an absolute magnitude $\mathrm{M_V\sim-5}$ and a bright/hot circumstellar shell made of amorphous carbon grains, RCBs are so distinctive that we should be able nowadays to find them everywhere in our Galaxy using publicly available catalogues. In the optical, the search is difficult as RCBs are known to undergo unpredictable photometric declines. However mono-epoch mid-infrared data can help us to discriminate RCBs among other dust-producing stars. The aim is to produce from the mid-infrared WISE and near-infrared 2MASS catalogues a new catalogue of reasonable size, enriched with RCB stars.}
{Colour-Colour cuts used on all stars detected are the main selection criteria. The selection efficiency was monitored using the 52 known RCBs located in the sky area covered by the WISE first preliminary data release.}
{It has been found that selection cuts in mid-infrared colour-colour diagrams are a very efficient method of discriminating RCBs from other stars. An RCB enriched catalogue made of only 1602 stars, with a high detection efficiency of about 77\%, was produced. Spectral energy distributions of 49 known RCBs and 5 known HdCs are also presented with estimates of their photosphere and circumstellar shell temperatures.}
{The newly released WISE all sky catalogue has proven to be a valuable resource in finding RCB stars. Actual scenarios predict that between 100 and 500 RCBs exist in our Galaxy. The newly created RCB enriched catalogue is an important step forward to significantly increase the number of known RCB stars and therefore better understand their origin.}


\authorrunning{Tisserand, P. }
\titlerunning{Tracking down R Coronae Borealis Stars from their mid-Infrared WISE colours}

\maketitle

\section{Introduction \label{sec_intro}}

There exists classes of stars that differ from the vast majority of the stars known in the sense that they are hydrogen-deficient. They are called extreme Helium stars (eHe), hydrogen-deficient Carbon stars (HdC) and R Coronae Borealis (RCB) stars. The last two are supergiant carbon-rich stars and have therefore strong spectroscopic similarity, but only RCBs are known to undergo unpredictable fast and large photometric declines (up to 9 mag over a few weeks) due to carbon clouds formed close to the line of sight that obscure the photosphere. Such particular events in such peculiar stars have made RCBs  much-followed objects among many generations of astronomers. Nowadays, RCBs have become even more interesting as they are increasingly suspected to result from the merger of two white dwarfs (one CO- and one He-), called the double degenerate (DD) scenario. The DD model has been strongly supported by the observations of an $^{18}$O overabundance in HdC and cool RCB stars \citep{2007ApJ...662.1220C,2010ApJ...714..144G}, and of surface abundance anomalies of a few elements, fluorine in particular \citep{2008ApJ...674.1068P,2011MNRAS.414.3599J}. RCBs are actually the favoured candidates to be the lower mass counterpart of Supernovae type Ia objects in a DD scenario \citep{2008ASPC..391..335F,2008arXiv0811.4646D}, therefore detailed studies of their peculiar and disparate atmosphere composition would help us to constrain simulations of merging events \citep{2011MNRAS.414.3599J}.

RCB stars are rare: we currently know about 50 of them in the Milky Way \citep[see][and references therein]{2008A&A...481..673T}, which is only a factor of two higher than in the Magellanic Clouds, where 23 RCBs are known \citep{2001ApJ...554..298A,2004A&A...424..245T,2009A&A...501..985T}. The small number of known RCB stars and the bias due to their discovery from different surveys prevent us from having a clear picture of their true spatial distribution. Different views are found in the literature. \citet{1985ApJS...58..661I} reported a scale height of $h\sim400$ pc assuming $\mathrm{M_{Bol} = -5}$ and concluded that RCBs are part of an old disk-like distribution. However, \citet{1998PASA...15..179C} noted that the Hipparcos velocity dispersion of RCB stars is similar to those of other cool Hydrogen-deficient carbon stars and extreme Helium stars, suggesting that RCB stars might have a bulge-like distribution. Recently, adding to the confusion, \citet{2008A&A...481..673T} found that the majority of Galactic RCB stars seem to be concentrated in the bulge with the surprising peculiarity of being distributed in a thin disk structure ($\mathrm{61<h^{RCB}_{bulge}<246}$ pc, 95\% c.l.). It is therefore necessary to increase the number of known RCB stars to constrain their spatial distribution and their age, but also understand their past evolution. I note that with an RCB phase lifetime of about $10^5$ years, as predicted by theoretical evolution models \citep{2002MNRAS.333..121S}, and an estimated He-CO white dwarfs merger birthrate between $\sim10^{-3}$ and $\sim5\times10^{-3}$ per year \citep{2001A&A...365..491N,2009ApJ...699.2026R}, we can expect between 100 and 500 RCB stars to exist in our Galaxy.

RCB stars are very bright, $\mathrm{-5\leqslant M_V\leqslant-3.5}$ \citep[Fig. 3]{2009A&A...501..985T}, and can therefore easily be found anywhere in our Galaxy. However, numerous observations are necessary to look for their main signature, large declines in luminosity. Well-sampled light curves, with a limiting magnitude of $\sim18$ for a large fraction of the sky, will not be available until the arrival of the LSST telescope \footnote{LSST: Large Synoptic Survey Telescope \citep{2008SerAJ.176....1I}}. Fortunately, RCBs are also known to possess relatively warm and bright circumstellar shells, easily detectable in the mid-infrared. These shells are made of amorphous carbon dust which translates to an almost featureless mid-infrared spectrum \citep{2011ApJ...739...37A}, unlike the spectra of classical old stars which have silicate and hydrogen rich dust shells. Therefore, one can imagine finding RCBs using only publicly available infrared catalogues. Such attempts have been made : \citet{2011A&A...529A.118T} found two new RCBs in the Galactic bulge using mainly mid-infrared Spitzer\footnote{Spitzer is a space telescope launched in 2003 to do infrared imaging and spectroscopy \citep{2004ApJS..154....1W}} GLIMPSE\footnote{GLIMPSE: Galactic Legacy Infrared Mid-Plane Survey Extraordinaire \citep{2009PASP..121..213C}} data and OGLE-III\footnote{OGLE-III: Optical Gravitational Lensing Experiment \citep{2003AcA....53..291U}} light curves, and \citet{2005ApJ...631L.147K} found cool objects in the Small Magellanic Cloud (SMC) with featureless spectra using the Spitzer spectrograph. Therefore it is interesting to look at how the first data release of the all-sky mid-infrared survey WISE (Wide-Field Infrared Survey Explorer) \citep{2010AJ....140.1868W} can improve our search for RCBs. This is the goal of the research described in this article.

In Section~\ref{sec_wise}, I describe briefly the WISE survey and catalogue and explain the impact of the [4.6] band bias observed on bright objects using the spectral energy distribution of known RCB stars. I describe in Sect.~\ref{sec_ana} all the criteria used in the analysis to select a small subsample of the WISE catalogue, enriched in RCB stars. Section~\ref{sec_discu} is a discussion of the outcome of the analysis and of the characteristics of the newly formed catalogue.

\section{WISE catalogue and the known RCB stars \label{sec_wise}}

\begin{figure}
\centering
\includegraphics[scale=0.4]{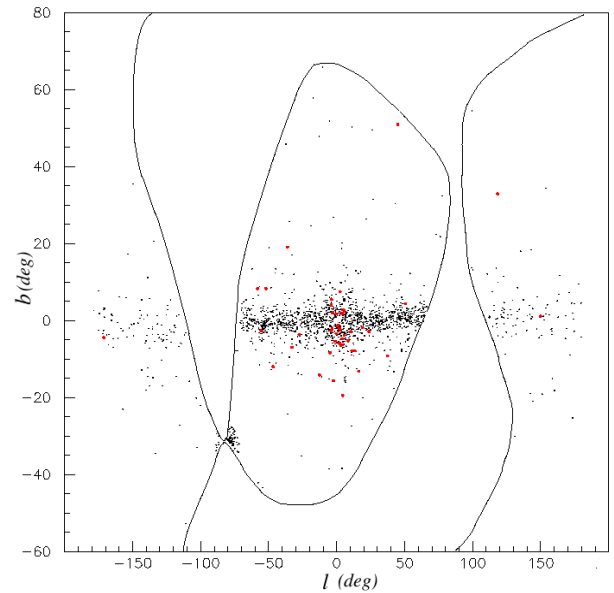}
\caption{Representation of the Galactic map, \textit{b} vs \textit{l}, with all objects selected by the analysis (black dots) and the known Galactic RCB stars (larger red dots). The black lines represent the approximate limits of the sky area released in the WISE-PDR1 catalogue. }
\label{map_lb}
\end{figure}

The Wide-field Infrared Survey Explorer (WISE) mapped the entire sky during 2010 in 3.4, 4.6, 12 and 22 $\mu$m with, respectively,  6.1, 6.4, 6.5 and 12.0 arcsec in angular resolution and 0.08, 0.11, 1.0 and 6.0 mJy in point source sensitivities at 5 sigma \citep{2010AJ....140.1868W}. A few months later, a preliminary data release (WISE-PDR1) was delivered from the first 105 days of observation. The catalogue contains 257 million sources spread over 57\% of the sky. Fortunately for an RCB star search, the released sky area covers the entire Galactic bulge (see Figure~\ref{map_lb}), where most of the known RCBs are located.

Indeed, 46 of 52 Galactic and 4 of 23 Magellanic RCBs are catalogued in WISE-PDR1, but also all 5 known HdCs and 4 DY Per type stars. All WISE magnitudes and the 1-sigma error associated are listed in Table~\ref{tab.WISE} for each object. In Table~\ref{tab.AKARI}, I also list fluxes and 1-sigma errors of all bright Galactic RCBs observed by the AKARI satellite, which did a mid-infrared all-sky survey in 2006 in 6 bands (centred at 9 and 18 $\mu$m with the IRC camera, and 65, 90, 140 and 160 $\mu$m with the FIS camera) \citep{2007PASJ...59S.369M}.

Photometry of bright non-saturated sources in WISE-PDR1 has an accuracy of about 2\% in [3.4], [4.6] and [12], and about 3\% for [22] \citep[see for more details][the WISE preliminary release explanatory supplement document]{2011wise.rept....1C}. The sensitivity varies significantly due to the different depth of coverage ($\sim10$ epochs on average), the background emission, and the source confusion. Saturation begins to affect sources brighter than approximately 8.0, 6.7, 3.8 and -0.4 mag respectively in all four bands. Most of the known RCBs are brighter than these limits, therefore only the PSF-fitting magnitudes will be used in the study. Measurements of saturated sources are realised with the non-saturated pixels in the objects'wings. Profile-fit photometry begins to fail for sources brighter than 1.0, 0.0, -2.0 and -6.0 mag, which are fortunately brighter than the brightest known RCB star. Photometric bias due to saturation remains small ($<0.1$ mag) for the [3.4], [12] and [22] WISE bands; however, an over-estimate in brightness is observed in the [4.6] band, up to nearly 1 mag for objects brighter than 3 mag above the saturation limit (which corresponds to $[4.6]\sim3.5$ mag). The impact of this bias on bright [4.6] objects will be discussed several times in this article, especially in Sect.~\ref{satbias}.

\subsection{WISE variability and classification flags\label{WISEflags}}

RCB shell brightness varies. \citet{1997MNRAS.285..317F} have shown that one can also observe in the L band (3.5 $\mu$m) the short-term oscillations that are observed in the optical ($\pm$0.5 mag), as the shell reflects light from the photosphere. However, the typical RCB large photometric declines are not observed in the mid-infrared as only the photosphere gets obscured by clouds. The effect of these clouds on the shell luminosity is different. Clouds produced in any direction, not just in the line of sight, will obscure the light coming from the star and therefore make the shell luminosity gradually fainter. Mid-infrared monitoring could therefore be used as a good indicator of change in the dust formation rate around the photosphere of an RCB star. A clear example is presented by \citet{1997MNRAS.285..317F} with UW Cen, where a variation of up to 3 mag in a $\sim1000$ days was observed. This phenomenon is also well described by \citet{1999ApJ...517L.143C}.

WISE-PDR1 gives a variability flag for each band. Therefore one can expect to get a positive variability flag for some of the RCBs catalogued. However, none of them was flagged to be variable, which is surprising as pulsations with periods between 30 to 60 days and amplitudes of $\sim0.5$ mag should have been noticed. The variability flag was calculated if an object was observed at least 11 times, but only 45\% of the RCBs have passed this threshold. (Z Umi has the record with 34 measurements, followed by R CrB and Y Mus with 20 measurements.) Furthermore, successive observations of a WISE field were made on a short time scale. They span at maximum a time range of a few days, which could be too short to observe variability in RCBs.

A classification (star/galaxy) flag is also available in WISE-PDR1, but it will not be used in further analysis as most of the catalogued RCBs were found to be not consistent with a point source.

\begin{table*}
\caption{The optical and near-infrared maximum apparent magnitudes determined from the AAVSO, ASAS, DENIS and 2MASS datasets. No interstellar extinction was applied to the magnitudes.
\label{tab.RCBsMAG}}
\medskip
\centering
\begin{tabular}{lccccccccc}
\hline
\hline
 Name & B & V & R & I & J & H & K & $\mathrm{E(B-V)}$ $^{a}$ & $\mathrm{A_V}$ $^{b}$ \\
\hline
\object{DY Cen} & 13.4 & 13.1 & 12.9 & 12.6 & 12.180 & 11.974 & 11.563 & 0.37 & 1.15 \\
\object{ES Aql} & 12.75 & 11.68 & -- & -- & -- & -- & -- & 0.32 & 0.97 \\
\object{FH Sct} & 13.6 & 12.2 & 11.5 & 10.7 & 9.212 & 8.549 & 7.585 & 0.82 (0.59)$^c$ & 2.52 (1.80)$^c$\\
\object{GU Sgr} & 11.7 & 10.35 & 9.8 & 9.326 & 8.545 & 8.269 & 7.993 & 0.49 & 1.52 \\
\object{MV Sgr} & 13.6 & 13.3 & 13.15 & 12.7 & 11.054 & 9.856 & 8.873 & 0.38 & 1.17 \\
\object{R CrB} & 6.6 & 5.8 & -- & -- & 5.364 & 5.089 & 4.564 & 0.03 & 0.11 \\
\object{RS Tel} & 10.66 & 9.9 & 9.35 & 9.12 & 8.707 & 8.449 & 7.915 & 0.09 & 0.27 \\
\object{RT Nor} & 11.25 & 10.2 & 9.6 & 9.2 & 8.684 & 8.457 & 8.171 & 0.24 & 0.73 \\
\object{RZ Nor} & 11.15 & 10.4 & 9.5 & 9.433 & 8.4 & 7.9 & 7.4 & 0.57 (0.24)$^c$ & 1.76 (0.73)$^c$\\
\object{RY Sgr} & 7.0 & 6.3 & 6.05 & 5.75 & 5.577 & 5.423 & 5.139 & 0.09 & 0.27 \\
\object{S Aps} & 11.0 & 9.7 & 9.0 & 8.25 & 7.269 & 6.844 & 6.412 & 0.14 & 0.42 \\
\object{SU Tau} & 10.8 & 9.8 & -- & -- & 7.60 & 7.05 & 6.5 & 0.73 & 2.24 \\
\object{SV Sge} & 12.25 & 10.45 & -- & 8.4 & 6.951 & 6.434 & 5.899 & 0.88 & 2.71 \\
\object{U Aqr} & 12.2 & 11.2 & -- & -- & 9.562 & 9.283 & 8.961 & 0.04 & 0.11 \\
\object{UV Cas} & 12.0 & 10.6 & 10.3 & 9.1 & 7.823 & 7.455 & 7.181 & 1.26 (0.76)$^c$ & 3.87 (2.35)$^c$\\
\object{UW Cen} & 10.0 & 9.4 & 8.8 & 8.5 & 8.0 & 7.4 & 6.6 & 0.39 & 1.19 \\
\object{UX Ant} & 12.6 & 12.0 & 11.7 & 11.5 & 11.148 & 10.976 & 10.695 & 0.09 & 0.29 \\
\object{V1157 Sgr} & -- & 11.5 & -- & -- & 8.497 & 7.843 & 7.083 & 0.14 & 0.43 \\
\object{V1783 Sgr} & -- & 10.6 & -- & 9.074 & 7.837 & 7.367 & 6.838 & 0.58 & 1.78 \\
\object{V2552 Oph} & -- & 10.9 & -- & -- & 8.686 & 8.387 & 8.163 & 0.82 & 2.52 \\
\object{V348 Sgr} & -- & 11.6 & -- & 11.353 & 10.253 & 8.824 & 7.260 & 0.34 & 1.03 \\
\object{V3795 Sgr} & -- & 10.95 & -- & 10.098 & 9.133 & 8.754 & 8.257 & 0.83 (0.29)$^c$ & 2.56 (0.91)$^c$\\
\object{V4017 Sgr} & -- & -- & -- & -- & 9.260 & 8.853 & 8.395 & 0.26 & 0.80 \\
\object{V482 Cyg} & 12.3 & 10.70 & 9.85 & 9.1 & 8.090 & 7.733 & 7.474 & 1.13 (0.70)$^c$ & 3.49 (2.17)$^c$\\
\object{V517 Oph} & -- & 11.4 & -- & -- & 7.507 & 6.803 & 6.104 & 0.71 & 2.19 \\
\object{V739 Sgr} & -- & 12.90 & -- & -- & 9.467 & 8.851 & 8.105 & 0.39 & 1.21 \\
\object{V854 Cen} & 7.6 & 7.05 & 6.85 & 6.55 & 6.106 & 5.695 & 4.875 & 0.12 & 0.36 \\
\object{V CrA} & -- & 9.7 & -- & -- & 8.700 & 8.307 & 7.491 & 0.11 & 0.34 \\
\object{VZ Sgr} & 11.1 & 10.4 & 9.9 & 9.5 & 9.013 & 8.743 & 8.243 & 0.31 & 0.95 \\
\object{WX CrA} & -- & 10.6 & -- & -- & 7.9 & 7.45 & 6.5 & 0.16 & 0.49 \\
\object{XX Cam} & 7.8 & 7.3 & 6.9 & 6.4 & 5.812 & 5.635 & 5.506 & 1.30 (0.25)$^c$ & 4.00 (0.76)$^c$\\
\object{Y Mus} & -- & 10.25 & -- & 9.55 & 8.602 & 8.406 & 8.243 & 0.83 (0.14)$^c$ & 2.56 (0.42)$^c$\\
\object{Z Umi} & 12.0 & 10.9 & 10.1 & 9.25 & 8.411 & 7.922 & 7.309 & 0.09 & 0.29 \\
\hline
\object{DY Per} & -- & 10.6 & -- & -- & 5.809 & 4.772 & 4.120 & 0.55 & 1.71 \\

\hline
\hline
\multicolumn{10}{l}{$^a$ From \citet{1998ApJ...500..525S} map, using 4 pixels interpolation. $^b$ Calculated using $\mathrm{E(B-V)/A_V=3.08}$ } \\
\multicolumn{10}{l}{$^c$ Values in parentheses are the reddening $\mathrm{E(B-V)}$ and the extinction $\mathrm{A_V}$ determined using the photosphere}\\
\multicolumn{10}{l}{ \hspace{2 mm} Effective temperatures found by \citet{2000A&A...353..287A}. The related stars are close to the Galactic plane ($\mathrm{\vert b \vert\leqslant5}$ deg).} \\
\end{tabular}
\end{table*}

\subsection{Spectral energy distributions}

The spectral energy distributions (SEDs) of all bright Galactic RCB and HdC stars were reconstructed using catalogues from 6 different surveys done in the last 10 years (see Figures \ref{SED1} to \ref{SED4}). In the optical, it is important to measure the maximum brightness of an RCB and only well sampled monitoring on a long time scale can give us a confident result. Therefore, light curves published by AAVSO\footnote{American Association of Variable Star Observers, URL: http://www.aavso.org/vstar/vsots/0100.shtml} (4 bands : B, V, R and I) and ASAS-3 \footnote{All Sky Automated Survey \citep{1997AcA....47..467P}, URL: http://www.astrouw.edu.pl/asas/?page=main} (V band) surveys have been analysed. The DENIS\footnote{DENIS:The Deep Near-Infrared Southern Sky Survey \citep{1994Ap&SS.217....3E}} I magnitude was also used if the epoch corresponded to a maximum brightness phase, as shown by AAVSO or ASAS-3 lightcurves. Overall, the maximum brightness measured in each optical band is accurate to $\sim0.05$ mag (1-sigma standard error). In the near-infrared, the 2MASS\footnote{2MASS: Two Micron All Sky Survey \citep{2006AJ....131.1163S}} J, H and K magnitudes were used. Due to the pulsating variability of RCB stars brightness, the accuracy of the maximum brightness measurement in these 3 bands is low, and estimated to be $\sim0.3$ mag. A carbon extinction correction was applied if the 2MASS epoch corresponded to a fading phase of an RCB. This was the case for only 7 stars : SU Tau, UW Cen, RZ Nor, V517 Oph, WX CrA, V348 Sgr and DY Per. We used for such corrections the $\Delta V$ magnitude variation observed at that particular epoch and the absorption coefficients of pure amorphous carbon dust presented by \citet[Fig. 2]{1995A&A...293..463G}. All magnitudes used in the study are listed in Table~\ref{tab.RCBsMAG} together with the interstellar reddening and extinction factors, $\mathrm{E(B-V)}$ and $\mathrm{A_V}$, obtained from the COBE/DIRBE\footnote{COBE: COsmic Background Explorer ; DIRBE: Diffuse Infrared Background Experiment} map \citep{1998ApJ...500..525S} with a 4 pixel interpolation. The 1-sigma error on E(B-V) is about 0.1 mag and becomes higher at higher extinction, up to 0.3 mag at $\mathrm{E(B-V)\sim1.2}$ mag \citep{2008A&A...481..673T}. These extinction coefficients were applied to all optical and near-infrared magnitudes using \citet{1992ApJ...395..130P} extinction curves, except for 7 stars located at low Galactic latitude ($\mathrm{\vert b\vert\leqslant5}$ deg) where \citet{1998ApJ...500..525S} note that the calculated reddening is uncertain and untrustworthy as no contaminating sources were removed\footnote{Effectively, for 5 of the 7 RCBs, the fitted photosphere effective temperatures were higher than 12000 K, when the \citet{1998ApJ...500..525S} map was used to correct for interstellar reddening. This is about 5000 K higher than expected.}. For these stars, the reddening was calculated from the \citet{2000A&A...353..287A} photosphere effective temperature determined using high resolution spectra, and the $\mathrm{V-I}$ colour index of the stars before reddening correction.

The SEDs were fitted using the program DUSTY \citep{2000ASPC..196...77N} with two (or three if necessary) black bodies, a shell made entirely of amorphous carbon grains and an MRN \citep{1977ApJ...217..425M} size distribution from 0.005 to 0.25 $\mu$m. The DUSTY models used have a photosphere temperature ranging from 2800 and 20000 K with a step of 100 K and a shell temperature ranging from 300 to 1200 K with a step of 50 K. The shell visual optical depth, $\mathrm{\tau_V}$, varied also between $10^{-3}$ and 10 on a logarithmic scale with 40 increments. The results are presented in Figures \ref{SED1} to \ref{SED4} together with the fitted effective temperatures. \citep[As mentioned earlier, for 7 RCBs located at low Galactic latitude, the photosphere effective temperature was fixed to the temperature determined by][]{2000A&A...353..287A}. In the mid-infrared, only three WISE magnitudes were used: the [4.6] band was not considered. AKARI magnitudes were fitted if no WISE magnitudes were available - this is the case with UV Cas, U Aqr and V482 Cyg - or if a clear third cold black body was observed (as with DY Cen, MV Sgr, UW Cen and WX CrA).

As described in Sect~\ref{WISEflags}, RCB' shell brightness varies. It is therefore not surprising to observe differences in luminosity between the WISE and AKARI data, which were taken 4 years apart. Such differences are visible on the SEDs presented, particularly with RT Nor, where a $\sim2$ mag variation is observed, but also with V1783 Sgr and V3795 Sgr. I note that the 2MASS K band can also be affected by shell brightness variation in the case of a hot shell (see WX CrA SED for example). The K flux was not used in the SED fit if it clearly contradicted the mid-infrared fluxes.

On average, the accuracy on the temperature and optical depth estimates is not better than $\sim10$\%, because of the combined effect of the interstellar dust extinction uncertainty (main effect), the missing U band magnitude and the fact that I did not use a model of a carbon rich photosphere\footnote{Carbon molecules CN and $C_2$ create strong absorption features in the optical and the near-infrared continuum.} \citep[see][]{1998A&A...330..659L}. I compared the photosphere effective temperature found to the more reliable one estimated by \citet{2000A&A...353..287A} for 17 RCBs using high resolution spectra \citep[temperatures from][are also indicated on the SED Figures]{2000A&A...353..287A} . Clearly, the SED method overestimates the photosphere effective temperature by about 600 K on average. On the other hand, for 4 of the 7 RCB stars (RZ Nor, Y MUs, UV Cas and V3795 Sgr) located at low Galactic latitude, the best SED model found with a photosphere temperature fixed to the one estimated by \citet{2000A&A...353..287A} and with an interstellar reddening deduced from this temperature, does not fit well the near-infrared magnitudes, indicating that the photosphere temperatures could be underestimated.

\begin{figure}
\centering
\includegraphics[scale=0.5]{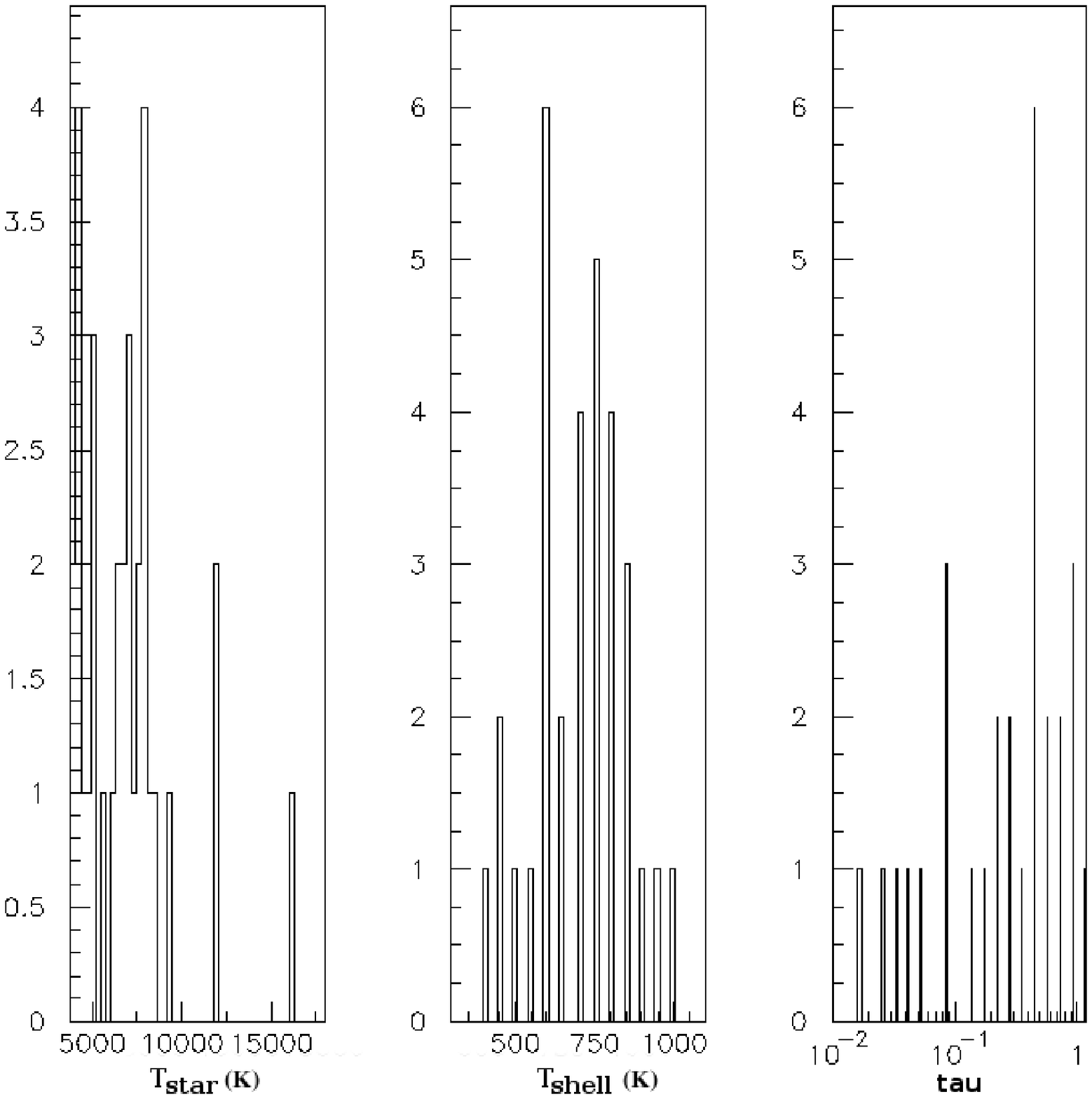}
\caption{From left to right, distribution of the fitted effective temperature of the RCBs' photosphere and shell, and the shell optical depth.}
\label{DistribTemp}
\end{figure}

Nevertheless, with this simple exercise, I can confirm that RCB stars have a wide range of photosphere temperatures, with the vast majority having a temperature between $\sim4000$ and 8000 K (spectral type K to F), with a few exceptions being hotter than 10000 K (DY Cen, MV Sgr and V348 Sgr) \citep{2002AJ....123.3387D}. RCBs also have a wide range of shell temperatures ranging from $\sim400$ to $\sim1000$ K. The typical RCB shell temperature is about 700 K with a visual optical depth of $\mathrm{\tau_V\sim0.4}$. The distributions of these three parameters are presented in Figure~\ref{DistribTemp}.

The SEDs of HdC stars, presented in Figure~\ref{SED4}, confirm that their spectral types are F-G, with an effective temperature ranging from 5500 to 7000 K. As no brightness declines have ever been recorded with HdC stars, it is not surprising not to detect a mid-infrared excess in their SEDs, but interestingly in the case of one HdC, HD 175893, one can fit a second blackbody which corresponds to a shell with an effective temperature of $\sim500$ K. HD 175893 could therefore be an RCB star going through a long period of low activity, corresponding to a low, if any, dust production rate. Furthermore, I note that HD 175893 has passed all the RCB selection criteria described below in Sect.~\ref{sec_ana}.

\subsubsection{WISE [4.6] band bias on bright objects\label{satbias}}

\begin{figure}
\centering
\includegraphics[scale=0.29]{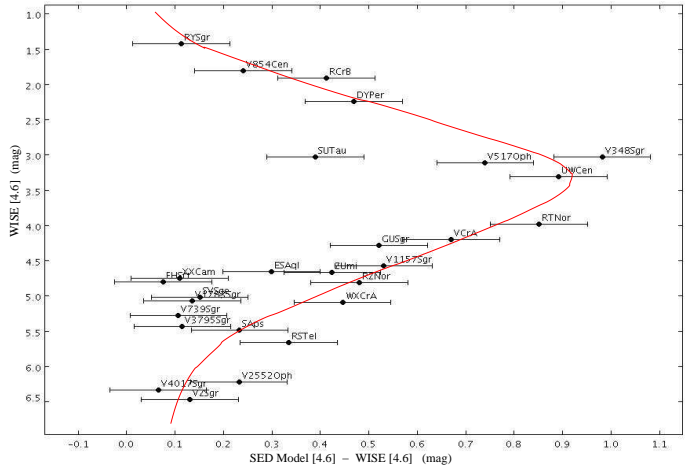}
\caption{Diagram representing the measured WISE [4.6] magnitudes of known RCB stars versus the difference in magnitude between the WISE measurement and the best SED model found. For objects brighter than $\mathrm{[4.6]\sim5.0}$, the [4.6] WISE brightness is overestimated, reaching $\sim0.9$ mag at $\mathrm{[4.6]\sim3.0}$ mag. The red line is a representation of this bias.}
\label{4.6ObsModel}
\end{figure}

An interesting output of this work comes from the difference of the [4.6] magnitude published by WISE and the one expected from the best SED model found. Figure~\ref{4.6ObsModel} summarises the situation. As mentioned earlier, the WISE team reported a bias for bright objects in the [4.6] band, whose brightness could be overestimated by almost 1 mag. I confirm this effect. One can see that for RCBs brighter than $\mathrm{[4.6]\sim5.0}$ mag, there exists a gradual increase of an overestimation of the [4.6] brightness, up to $\sim0.9$ mag, and a decrease after $\mathrm{[4.6]\sim3.0}$ mag down to the nominal brightness. RCBs whose [4.6] magnitude is certainly affected (with $\mathrm{5.0\leqslant [4.6] \leqslant 1.5}$ mag) are indicated in Table~\ref{tab.WISE}. The WISE team has not yet found the source of this bias. This significant effect will be taken into account in the analysis described below.

\section{Analysis \label{sec_ana}}

Using the WISE and 2MASS catalogues, a series of pragmatic selection criteria was developed and optimised to create a reasonable size catalogue enriched in RCB stars located within 50 kpc. The criteria are described below. The detection efficiency of each cut was controlled using 52 known RCBs (see Table~\ref{tab.Selection}).

Colour-Colour cuts on all stars detected in every four WISE and three 2MASS broadbands are the main selection criteria used, as they do not impact on the distance of the RCB. These selection criteria are as follows.

\begin{enumerate}
\item The first selection criterion is simply to keep all objects detected and listed in the WISE catalogue. This process is made difficult by the brightness of RCB stars (often saturated) but also by the blending effect at low Galactic latitude where many RCB stars are expected to be. 52 known RCBs are expected to have been catalogued in WISE-PDR1 from the sky area released (see Figure~\ref{map_lb}). However, two are missing, OGLE-GC-RCB-1 \& -2 which were discovered already from their shell brightness and colours using the Spitzer-GLIMPSE catalogue \citep{2011A&A...529A.118T}. These 2 RCBs are located at very low Galactic latitude ($b\sim3.0$ and $\sim-1.8$ deg) in a very crowded field. They are clearly bright and distinguishable on WISE reference images, in all bands, but the WISE source finder did not succeed in retrieving them. This first criterion has therefore an important effect on the overall analysis and final outcome completeness. However, this issue may be corrected in further WISE data releases.

\item The objects selected have to have an entry in all of the four WISE bands. The main WISE criterion to accept an object in the catalogue is that this object should have been detected with a signal-to-noise higher than 7 in at least 1 band and detected in 4 frames minimum. At high Galactic latitude, the detected source distribution remains relatively uniform up to $\mathrm{\left[3.0\right]<16.5}$, $\mathrm{\left[4.6\right]<16.0}$, $\mathrm{\left[12\right]<12.5}$ and $\mathrm{\left[22\right]<9.0}$ mag. The fact that the faintest Magellanic RCB, EROS2-LMC-RCB-5, detected in WISE-PDR1 has magnitudes of $\sim11.9$, $\sim11.4$, $\sim8.6$ and $\sim7.4$ mag respectively in all four WISE bands, gives us strong confidence that all RCBs located within $\sim50$ kpc could be detected. RY Sgr is the only RCB star that did not pass this criterion. The problem is related to the other end of the brightness source distribution, as RY Sgr is the second brightest RCB star known in the optical after R CrB, but the brightest in the mid-infrared. RY Sgr failed to have an entry in the $\left[12\right]$ band. The impact on the overall detection efficiency is minimal, as the vast majority of the RCBs we are looking for are fainter.

\begin{figure}
\centering
\includegraphics[scale=0.5]{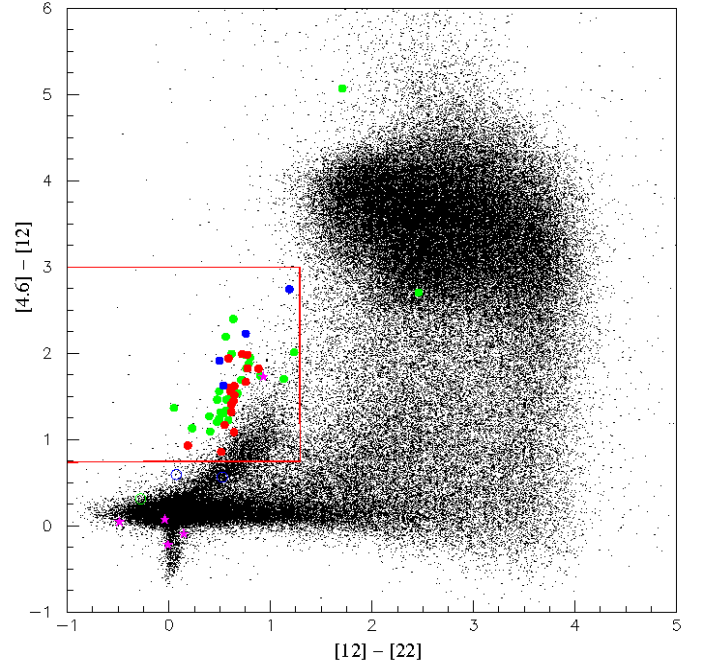}
\caption{Diagram of [4.6]-[12] vs [12]-[22]. The black dots represent 3\% of all objects that have been catalogued in all the WISE and 2MASS bands. Larger green dots represent bright known Galactic RCB stars (the two outliers are discussed in the text); red dots correspond to the Galactic RCBs found toward the bulge (with a majority being located inside the bulge) \citep{2008A&A...481..673T, 2005AJ....130.2293Z}; blue dots are 4 RCB stars located in the southern-east area of the Large Magellanic Cloud. The purple stars correspond to the 5 known HdC stars and the circles to DYPer type stars with a colour coding identical to RCB stars. The red lines represent the selection cuts used in the analysis.}
\label{M4M12M22big}
\end{figure}

\item All selected objects have to have an entry in all J, H and K bands of the 2MASS catalogue. All 49 remaining known RCBs have passed this selection. The 2MASS point source magnitude limits are $\sim15.8$, $\sim15.1$ and $\sim14.3$ mag respectively in J, H and K. With $\mathrm{M_K\sim-6}$, an RCB star located within $\sim50$ kpc should have an entry in the 2MASS catalogue if there is not a very strong interstellar extinction on the line of sight ($\mathrm{A^{inter.}_V\lesssim 15}$ mag). Also, the maximum extinction due to a cloud of carbon soot is known to be $\Delta\sim9$ mag in V \citep{1996PASP..108..225C}, which corresponds to $\sim3$ mag in J and $\sim1.6$ mag in K. Even in that extreme scenario, an RCB star located within $\sim50$ kpc should have been detected by the 2MASS survey under reasonable interstellar extinction conditions ($\mathrm{A^{clouds}_V\lesssim4}$ mag).

\item The first Colour-Colour selection cut was applied on the WISE catalogue to select a reasonable number of objects to work with : 
\begin{equation}
$$ 0.75<[4.6]-[12]<3.0 \hspace{3 mm}\&\hspace{3 mm} [12]-[22]<1.3 $$.
\label{eq.cut1}
\end{equation}

The selection is illustrated in Figure~\ref{M4M12M22big}. One can see that this selection is very effective as it selects all objects with a clear shell signal and eliminates most of the galaxies located in the top-right corner of the diagram. It keeps only 5\% of all objects catalogued, but rejects only 2 of the RCB stars used as references. These 2 RCBs are DY Cen and MV Sgr. They do not resemble the majority of RCB stars known, as they are hot ($\mathrm{T_{eff}\sim12000}$ K, see Figure~\ref{SED1}) and surrounded with multiple shells. Furthermore, and more interestingly, RCB stars are located in an underpopulated area of the diagram. They lie in an area well separated from the vast majority of the cooler AGB stars. This selection is refined in selection 6, and a discussion of the impact of the bias that affects bright objects in [4.6] is also presented.

\item RCB stars are known to present an excess in the near-infrared compared to classical supergiant stars \citep[see][Fig.4]{2009A&A...501..985T} due to the warm circumstellar shell that contributes significantly to the near-infrared fluxes. A selection has therefore been applied in the $\mathrm{J-H}$ vs. $\mathrm{H-K}$ diagram. Because the interstellar extinction affects these magnitudes significantly, cuts were developed for the 3 following Galactic latitude ranges, A, B and C: A: $b>2$ deg, B: $\mathrm{1\leqslant b < 2}$ deg (which corresponds to $\mathrm{\left\langle A_V \right\rangle \sim2}$ mag) and C: $\mathrm{b\leqslant 1}$ deg (with $\mathrm{\left\langle A_V \right\rangle \sim5}$ mag). Three selection cuts were applied and are described below. The parameters for each Galactic latitude range are given in square brackets as follows [A,B,C]:

\begin{eqnarray}
&& \hspace{5 mm}(H-K)>[0.2,0.32,0.5] \nonumber \\
&& \mathrm{if}\hspace{1 mm}(H-K)\leqslant[0.8,0.92,1.1] : \nonumber \\
&& \hspace{5 mm}(J-H)<(H-K)+[0.2,0.28,0.4] \nonumber \\
&& \mathrm{if}\hspace{1 mm}(H-K)>[0.8,0.92,1.1] : \nonumber \\
&& \hspace{5 mm}(J-H)<(5/3)(H-K)-1/3
\label{eq.cut2}
\end{eqnarray}

The interstellar extinction vector corresponds to $\mathrm{(J-H)/(H-K)\sim5/3}$, therefore the last selection cut is identical for each Galactic latitude range. These 3 selection cuts are represented on Figure~\ref{JHK_M4M12M22}. Only 2\% of the remaining objects passed this criterion, and only two RCB stars, Y Mus and XX Cam, were eliminated. 

\begin{figure}
\centering
\includegraphics[scale=0.5]{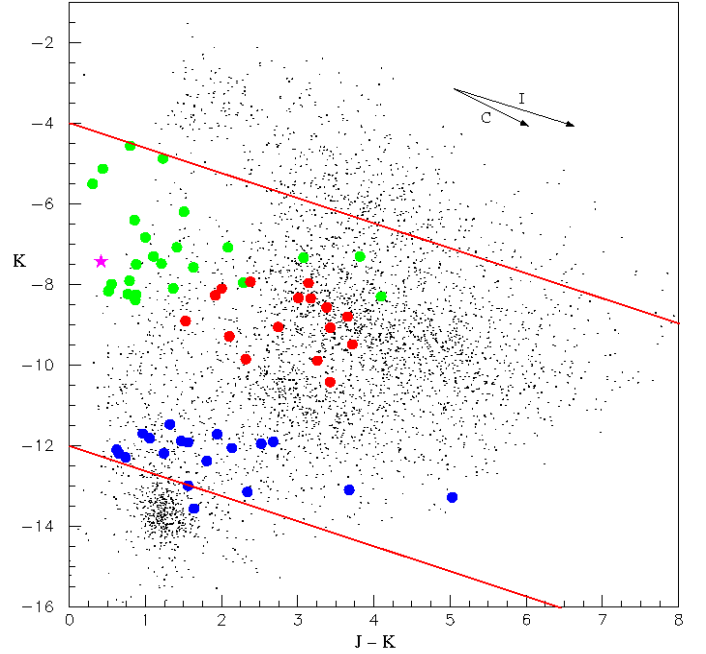}
\caption{Colour-magnitude diagram K vs $\mathrm{ J-K}$. Black points represent the remaining objects after application of selection criterion 6. The large dots correspond to RCB or HdC stars with the same colour coding as in Figure~\ref{M4M12M22big}. Extinction due to amorphous carbon grains (C) and interstellar dust (I) is represented with two vectors. The red lines represent selection cut number 7.}
\label{K_JK}
\end{figure}

\item The selection criterion 4 is readjusted to reject most of the AGB stars, and the applied cut is illustrated in Figure~\ref{JHK_M4M12M22} :

\begin{equation}
$$ [4.6]-[12] > 1.8([12]-[22]) + 0.03$$
\label{eq.cut3}
\end{equation}

As discussed in Sect.~\ref{satbias}, the brightness of bright objects in [4.6] is overestimated. RCBs affected by this bias are indicated in Table~\ref{tab.WISE}, and the estimated corrections are represented in Figure~\ref{JHK_M4M12M22} with black vertical lines. After correction, one can see that none of the RCBs reached the upper limit set at 3.0 mag on $\mathrm{[4.6]-[12]}$. The impact of the [4.6] bias on this selection criteria is therefore very limited. On the other hand, two out of the three RCBs that are eliminated by the present selection would have been selected if the bias did not exist. As we can expect that many of the RCBs we are looking for are relatively bright in the [4.6] band, all 3 known RCBs rejected will nevertheless be counted in the detection efficiency estimation.

\item The present criterion target RCB brightness and therefore their distance modulus. A selection was applied on the colour - magnitude diagram K vs $\mathrm{J-K}$. The colour $\mathrm{J-K}$ of RCBs is due to three effects: the interstellar reddening, the warm circumstellar shell contribution to near-infrared fluxes and for some RCBs the reddening due to clouds made of carbon soot during an extinction event (at maximum $\mathrm{\Delta(J-K)_{cloud}\sim 1.4}$ mag). The RCBs' $\mathrm{J-K}$ colour is therefore not simple and the selection limits slope were chosen with care. They are illustrated on Figure~\ref{K_JK}. The goal of this analysis is to create a catalogue enriched with RCB stars located within 50 kpc, therefore all RCBs discovered in the Large Magellanic Cloud (LMC) were represented in this diagram (they all have an entry in the 2MASS catalogue). They were used to estimate the lower end of the brightness cut. Only 2, EROS2-LMC-RCB-2 \& -3, out of 19 LMC RCBs were not selected.
\begin{equation}
$$ (5/8)(J-K)+4 \hspace{1 mm}< K < \hspace{1 mm}(5/8)(J-K)+12 $$
\label{eq.cut4}
\end{equation}
On the bright end of the selection, 2 of the brightest known RCBs, R CrB and V854 Cen, are located close to the upper limit. Very bright RCBs are not the primary target of this analysis as very few are expected, therefore only a small margin was used in this choice.

\begin{figure*}
\centering
\includegraphics[scale=0.45]{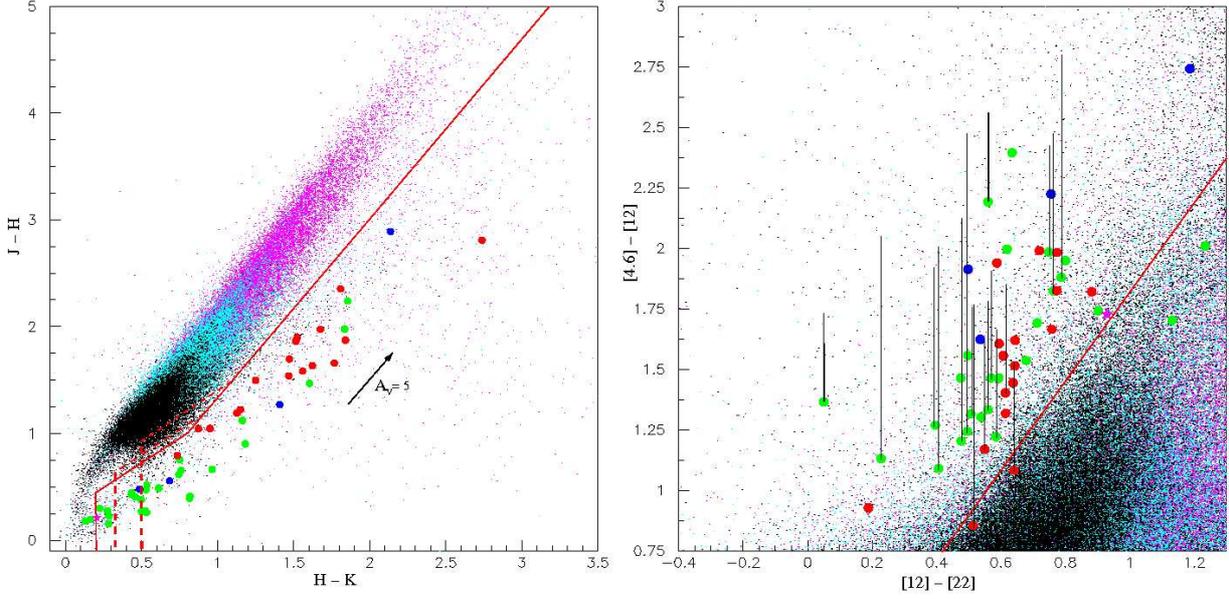}
\caption{In both figures, all points represent objects that have passed the first 4 selection criteria. They are colour coded to represent their Galactic latitude \textit{b} and therefore the interstellar reddening that affects their near-infrared J, H and K magnitudes: black points correspond to objects with $\textit{b}>2$ deg, light blue ones are objects with $\mathrm{1\leqslant\textit{b}\leqslant2}$ deg (corresponding to average extinction of $\mathrm{\left\langle A_V \right\rangle \sim2}$ mag), and purple ones are objects located close to the Galactic plane, with $\mathrm{\textit{b}<1}$ deg (with $\mathrm{\left\langle A_V \right\rangle \sim5}$ mag). The larger dots and stars represent known RCB and HdC stars with identical colour coding to Figure~\ref{M4M12M22big}. The dashed and solid red lines represent the selection cuts number 5 (left) and 6 (right) used in the analysis. Left: $\mathrm{J-H}$ vs $\mathrm{H-K}$ diagram. Right: $\mathrm{[4.6]-[12}$ vs $\mathrm{[12]-[22]}$ diagram. The black vertical lines correspond to the brightness correction that needs to be applied to bright RCBs in the [4.6] band due to the photometric bias (see Sect.~\ref{satbias}). }
\label{JHK_M4M12M22}
\end{figure*}

\begin{figure*}
\centering
\includegraphics[scale=0.45]{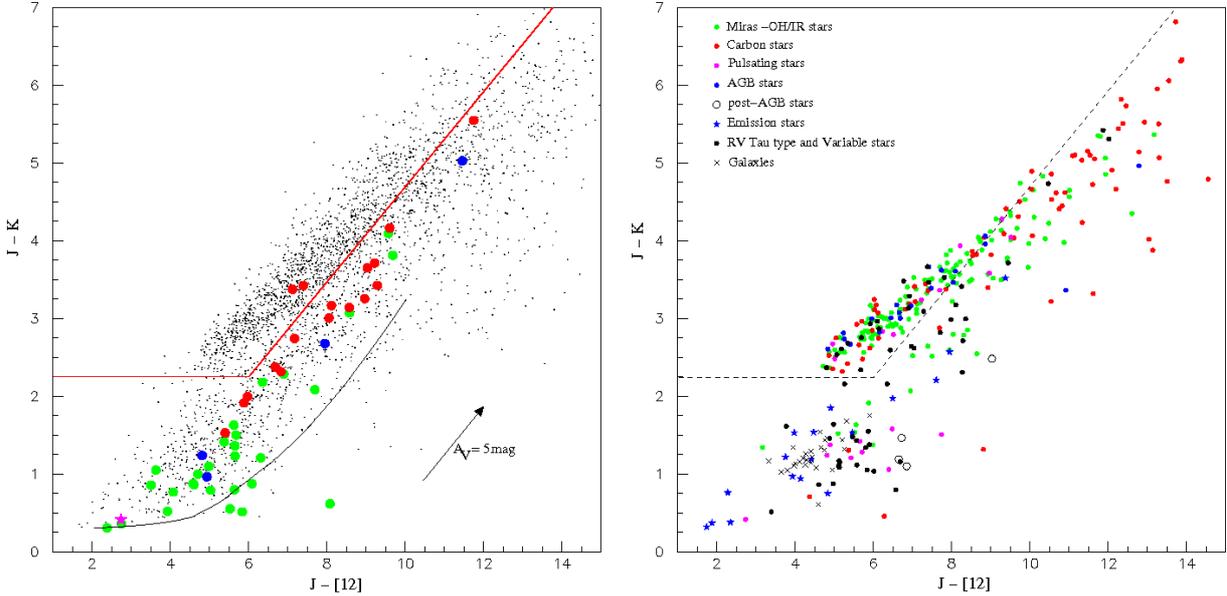}
\caption{Both diagrams represent $\mathrm{J-K}$ versus $\mathrm{J-[12]}$. Left: the black dots represent the remaining objects after application of selection criterion 7. The red line represents the selection cut number 8. The larger dots correspond to RCB or HdC stars that have also passed the first 7 criteria, with the same colour coding as in Figure~\ref{M4M12M22big}. The vector represents the interstellar reddening. The black curve corresponds to the combination of blackbodies consisting of a 6000 K star and a 700 K shell made of amorphous dust grains, in various proportions ranging from all "star" to all "shell".  Right: the symbols represent the classification found in SIMBAD. About 15\% of the objects selected with the first seven criteria have an entry in the SIMBAD database with a matching radius of 5 arcsec. The dashed line represents the selection cut number 8.}
\label{JK_J12}
\end{figure*}

\begin{figure*}
\centering
\includegraphics[scale=0.45]{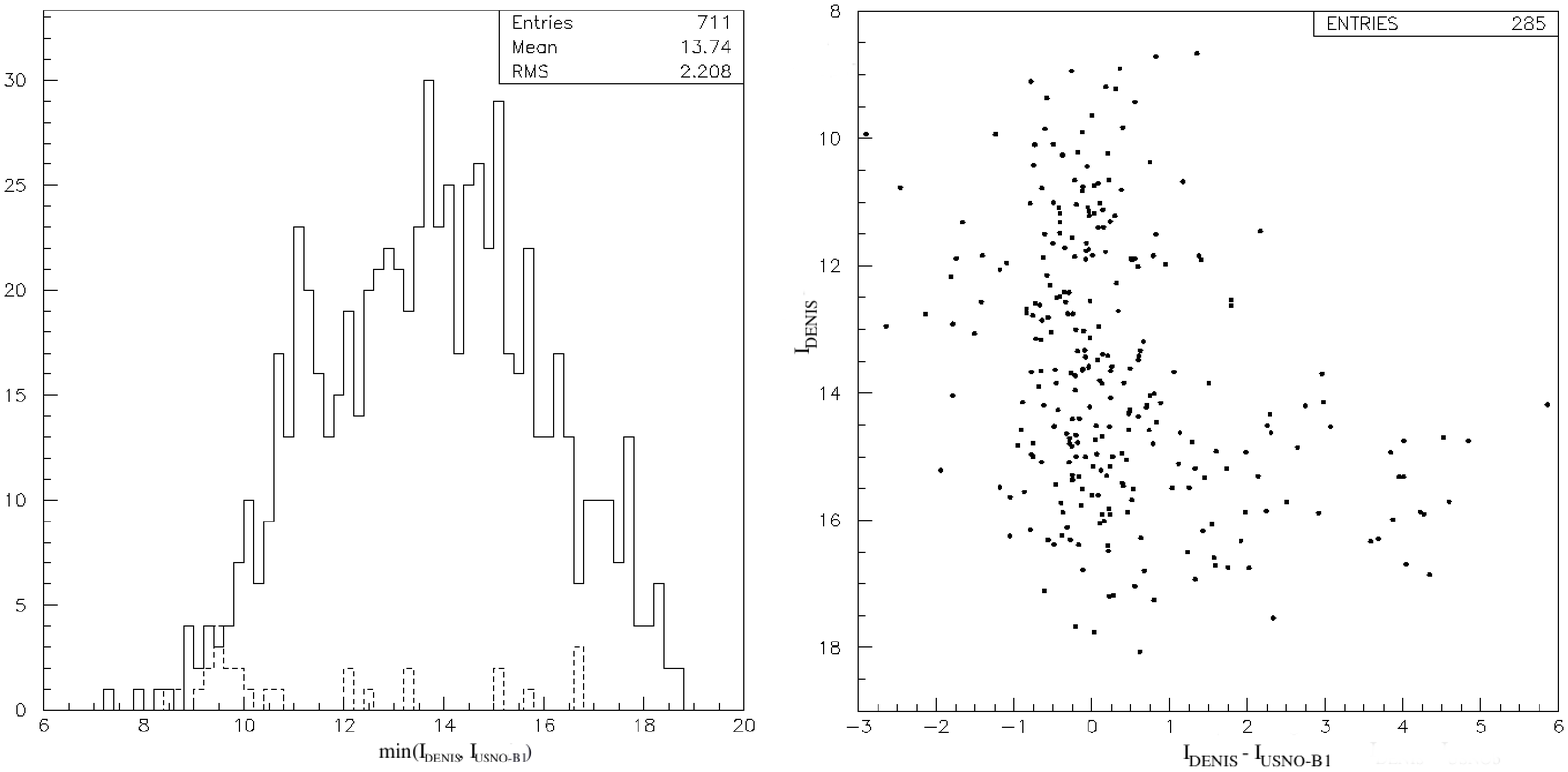}
\caption{Left: distribution of the I magnitude for 711 of 1602 objects selected by the analysis. Only 711 objects have a valid I band magnitude in the DENIS and/or USNO-B1 catalogues. The dotted lines represent the same distribution for known RCB stars. Right: $\mathrm{I_{DENIS}}$ vs $\mathrm{I_{DENIS}-I_{USNO-B1}}$ for the 285 of 1602 objects that have a valid I magnitude in both catalogues. The difference in magnitude between both catalogues covers about 9 magnitudes, showing that many selected objects undergo large photometric variation. }
\label{Denis_USNO_I}
\end{figure*}

\item As mentioned earlier, the RCB's warm circumstellar shell can contribute significantly to the near-infrared fluxes. This effect can be understood in a $\mathrm{J-K}$ vs $\mathrm{J-[12]}$ Colour-Colour diagram. Effectively, a correlation between both colours is observed for RCBs (see Fig.~\ref{JK_J12}). When the RCB shell becomes brighter, its flux contribution to the K band increases and as a consequence $\mathrm{J-K}$ becomes redder. This effect is in combination with the interstellar reddening effect. Interestingly, it is also relatively simple in that diagram to reject the remaining main contaminants, the Mira type stars. Indeed, they are located in an isolated area of the diagram. The applied cuts were designed to optimise the rejection of Miras without impacting too much on the RCB detection efficiency. The second selection follows the interstellar reddening vector:

\begin{equation}
$$ J-K\leqslant2.25 \hspace{2 mm} \mathrm{if}\hspace{2 mm} (J-[12])<6$$
\end{equation}

\begin{equation}
$$ J-K\leqslant0.61 (J-[12])-1.43 \hspace{2 mm}\mathrm{if}\hspace{2 mm} (J-[12])\geqslant6$$
\label{eq.cut5}
\end{equation}

Only 2 known RCB stars, out of the 42 that passed all the precedent criteria, were rejected by the present selection. They are EROS2-RCB-CG-12 and MACHO-308-38099.66. It is interesting to note that the former one has a light curve that presents large oscillations and may therefore be a Mira type star, instead of an RCB \citep[see][Fig.9]{2008A&A...481..673T}.

\end{enumerate}

\begin{table*} 
\caption{Number of selected objects after each selection criterion.
\label{tab.Selection}}
\medskip
\centering
\begin{tabular}{lrcl}
\hline
\hline
Selection criterion & Number of & Number of & RCB stars   \\
 & WISE objects & known RCBs & eliminated\\
\hline
0: Located in WISE-PDR1 sky area  &  & 52 & \\
1: Catalogued by WISE  & 257310278 & 50 & OGLE-GC-RCB-1 \& -2\\
  \hspace{3 mm} Catalogued in the [22] band & 12192351 & 50 & \\
2: Catalogued in all four WISE bands & 9636678 & 49 & RY Sgr\\
3: Catalogued in all three 2MASS bands & 6832051 &49 & \\
4: Cut on ($\mathrm{[4.6]-[12]}$ vs $\mathrm{[12]-[22]}$) & 337359 & 47 & DY Cen and MV Sgr \\
5: Cut on ($\mathrm{J-H}$ vs $\mathrm{H-K}$) & 9660 & 45 & Y Mus and XX Cam\\
6: Cut on ($\mathrm{[4.6]-[12]}$ vs $\mathrm{[12]-[22]}$) & 4146 & 42 & EROS2-RCB-CG-3 \& -5 and SV Sge\\
7: Cut on (K vs $\mathrm{J-K}$) & 3058 & 42 & \\
8: Cut on ($\mathrm{J-K}$ vs $\mathrm{J-[12]}$) & 1643 & 40 & EROS2-RCB-CG-12 and MACHO-308-38099.66\\
\hline
\hspace{2 mm}Final (without the known RCBs + HD 175893) & 1602 & - & \\
\hline
\end{tabular}
\end{table*}


\section{Discussion \label{sec_discu}}

1602 objects have passed the pragmatic selection criteria that have just been enumerated. Of the 52 already known RCBs that are located inside the sky area covered by the WISE survey preliminary data release and used as reference, 41 have passed all 8 criteria, which corresponds to a detection efficiency of about 77\%. For the detection efficiency calculation, it would have been preferable to use a Monte-Carlo simulation of RCBs with a range of photosphere and shell temperatures, but our knowledge of the distribution of these last parameters is very limited. It is worth noting that the RCB sample used as reference is biased toward bright RCBs and therefore had a negative impact on the detection efficiency (see criteria 2 and 6). Bright RCBs are not the primary target of this analysis as very few of them, if any, are expected to have not been discovered yet. At the other end, the selection is biased against RCB stars with faint shells and RCB stars in a crowded environment (see criteria 1 and 5). In the last case, a detection algorithm correction in a future WISE data release may resolve the issue.

About 70\% of the objects selected are located towards the Galactic bulge (see Figure~\ref{map_lb}), at $\pm$5 deg from the Galactic plane, and 85 ($\sim5$\%) are located in the Large Magellanic Cloud, whose sky area has been only partially covered in WISE-PDR1.

The final catalogue was cross-matched with the DENIS and USNO-B1\footnote{USNO: United States Naval Observatory \citep{2003AJ....125..984M}} catalogues, using a strict matching radius of 1 arcsec. Only 895 objects have an entry in the DENIS catalogue and 538 in the USNO-B1 one. DENIS did a survey of all the southern sky in I, J and K bands, while USNO-B1 covers the entire sky in B, R and I (2 epochs in B and R). Overall, using both catalogues, 711 objects have a valid I band magnitude, but surprisingly only 285 have a valid entry in both catalogues. The I band magnitude distribution is presented in Figure~\ref{Denis_USNO_I}, as well as the difference in magnitude between $\mathrm{I_{DENIS}}$ and $\mathrm{I_{USNO-B1}}$. With an absolute magnitude $\mathrm{M_I\sim-5}$ mag \citep{2009A&A...501..985T}, an RCB star located at 50 kpc would have an apparent magnitude of $I\sim13.5$ mag during a maximum brightness phase, which is brighter than the median ($\sim$14) of the distribution presented. The $\mathrm{I_{DENIS}}$ and $\mathrm{I_{USNO-B1}}$ magnitudes should be used to prioritise further follow-up. During some large photometric decline phases, some RCB stars could have been observed at fainter magnitude up to the magnitude limit ($\mathrm{I_{DENIS,lim}\sim18.5}$ mag, or $\sim17$ mag in crowded area): Figure~\ref{Denis_USNO_I}, on the right, shows that some of the selected objects have known photometric variation up to 6 mags between both survey epochs.

The fact that only $\sim33$\% of the 1602 selected objects have an entry in the USNO-B1 catalogue indicates that most of those 1602 objects are faint, with a visual magnitude lower than 20. The interstellar extinction has certainly an influence on this low percentage, but it is not the absolute answer as half of these faint objects are located at 2 degrees or more from the galactic plane ($\mathrm{\mid\textit{b}\mid >2}$ deg) where extinction $\mathrm{A_V}$ is lower than $\sim2$ mag. Under such extinction, an RCB star at maximum brightness and located within 50 kpc is still detectable, therefore these objects could be intrinsically optically faint and a contaminant of our research. However, they may also be (or a fraction of them) RCB stars that are heavily dust enshrouded. If one doesn't expect that during the lifetime of an RCB star, there exists a long phase of high dust production rate, then these objects should be considered as weak candidates. However, it is worth noting that the RCB star V854 Cen remained 7 mag below its maximum brightness for about 50 years \citep{1986IAUC.4245....1G}.

I note also that a third of the 1602 selected objects have a WISE $\mathrm{[4.6]}$ magnitude brighter than 5 mag. The $\mathrm{[4.6]}$ brightness of these objects is therefore overestimated (see Sect.~\ref{satbias}).

The selected list of objects was also cross-matched with the SIMBAD database with a matching radius of 5 arcsec. Figure~\ref{JK_J12}, right side, summarises the situation.  569 of the 1602 objects were found to have a classification, but 307 of those are classified only as IRAS sources and 30 others only as stars, without further informations. However, of the 232 remaining, 62 stars are classified interestingly as Carbon stars; they are stars that will need to be looked at closely in future photometric and spectroscopic follow-ups. Also, 46 objects are classified as Variable, Pulsating, Semi-regular or RV Tauri stars. RCB stars are known to present periodic pulsating variability with an amplitude of $\sim0.5$ mag; therefore these stars will be interesting to monitor and study more closely. Despite the fact that many of the objects classified as Mira or OH/IR stars were rejected using selection criterion 8, 49 remained in the final sample. Most of them are certainly genuine, particularly at $\mathrm{J-[12]>9}$ mag, but it is also worth indicating that a few Miras are located in the $\mathrm{J-K}$ vs $\mathrm{J-[12]}$ diagram at a position not expected for these type of object and can therefore be suspected of misclassification. There are 14 such objects with $\mathrm{J-K<3}$ mag. A visual inspection of the ASAS-3 lightcurves of these objects have shown that 2 of them present Mira-type photometric variation (Id 45 and 1197), 2 remain stable (Id 623 and 1409), but more interestingly 3 present clear photometric variations typical of RCB stars: Id 683, 793 and 917, which are named respectively \object{V653 Sco}, \object{IO Nor} and \object{V581 CrA}. The remaining 7 objects have not been catalogued and followed-up by the ASAS-3 survey. I note that 26 objects are classified as galaxies or clusters of galaxies. These objects are faint, with $K>12$ mag, indicating that galaxies could therefore be a major contaminant below this limit (270 on the 1602 selected objects are fainter than $K>12$). Finally, 17 objects are classified as Emission stars; such objects could be contaminants in our search for RCB stars, but it is interesting to note that hot RCBs, such as DY Cen, MV Sgr and V348 Sgr, present also an emission type spectrum \citep{2002AJ....123.3387D}. Furthermore, during faint phases, RCBs spectrum generally presents many emission lines \citep{1996PASP..108..225C}.

There does not exist any reference sample of stars to estimate the fraction of RCB stars that are potentially in the catalogue. Furthermore, little is known on the different phases during the lifetime of an RCB star and as mentioned earlier, a significant fraction of RCB stars could be heavily dust-enshrouded and therefore faint. However one can use the objects that have been classified in the SIMBAD database to obtain a crude estimate of this fraction: $\sim15$\% of these objects are RCBs, it would then corresponds to about 240 new RCBs.



The RCB enriched catalogue is available from the following URL:  http://www.mso.anu.edu.au/$\sim$tisseran/RCB/ and will also be available through the VizieR\footnote{URL: http://vizier.u-strasbg.fr/viz-bin/VizieR} catalogue service. A short version of it is given as example in Table~\ref{tab.short_version}. The equatorial and Galactic coordinates as well as all four WISE, three 2MASS, three DENIS magnitudes and their 1-sigma errors are listed for all 1602 objects selected. The five USNO-B1 magnitudes are also listed, but not the individual measurement error as they were not delivered in the original catalogue \citep[see][for an estimate of the photometric accuracy]{2003AJ....125..984M}. If one magnitude was not available, its value was replaced by the number -99. Also, if more than one epoch were available in the DENIS or USNO-B1 catalogues for one particular object, only the epoch related to the brightest magnitudes was kept. The last column of the catalogue gives the SIMBAD classification\footnote{URL of the object types in SIMBAD (classification version : 12-Jul-2011): http://simbad.u-strasbg.fr/simbad/sim-display?data=otypes}, as of July 2011, found using a 5 arcsec matching radius. An underscore character was given to the objects that had no classification in SIMBAD.

\section{Conclusion\label{sec_concl}}

Using both the 2MASS and WISE preliminary data release catalogues, I have created a catalogue enriched with Galactic RCB stars that lie within a distance $\sim50$ kpc. The selection criteria to select a subsample of objects with mid- and near- infrared properties similar to RCBs, were mostly based on colour-colour diagrams. This catalogue contains only 1602 entries, with $\sim70$\% of the selected objects being located toward the Galactic bulge. A reference sample of 52 known RCB stars was used to monitor the detection efficiency of the selection. About 77\% of them were recovered. Such a high detection efficiency gives strong support to the RCB content of this catalogue. Each selected object will now need to be followed-up spectroscopically to discover its true nature. It is encouraging to observe that 3 of these 1602 selected objects, Id 683, 793 and 917 (resp. \object{V653 Sco}, \object{IO Nor} and \object{V581 CrA}) were found to present ASAS-3 lightcurves with brightness variations typical of RCB stars, but are misclassified in the SIMBAD database as Miras. Also, \citet{2011PASP..123.1149K} has recently discovered that NSV 11154 is a new RCB stars. This star is listed in the catalogue under the Id 1240.

An analysis of the Spectral Energy Distribution of known RCB stars confirms that RCBs effective temperatures range mostly between 4000 and 8000 K (spectral type K to F), with a few exceptions being hotter than 10000 K. The RCB shell effective temperatures range between 400 and 1000 K. The typical shell temperature being $\sim700$ K with a visual optical depth of $\tau_V\sim0.4$.
HD 175893 is the only HdC star (spectral type G to F) that presents an infrared excess indicating the existence of a warm circumstellar shell. Furthermore, HD 175893 has passed all the selection criteria to create the RCB enriched catalogue, therefore HD 175893 should be considered to be an RCB star under a phase of low activity in terms of dust production.

\begin{figure*}
\centering
\includegraphics[scale=0.9]{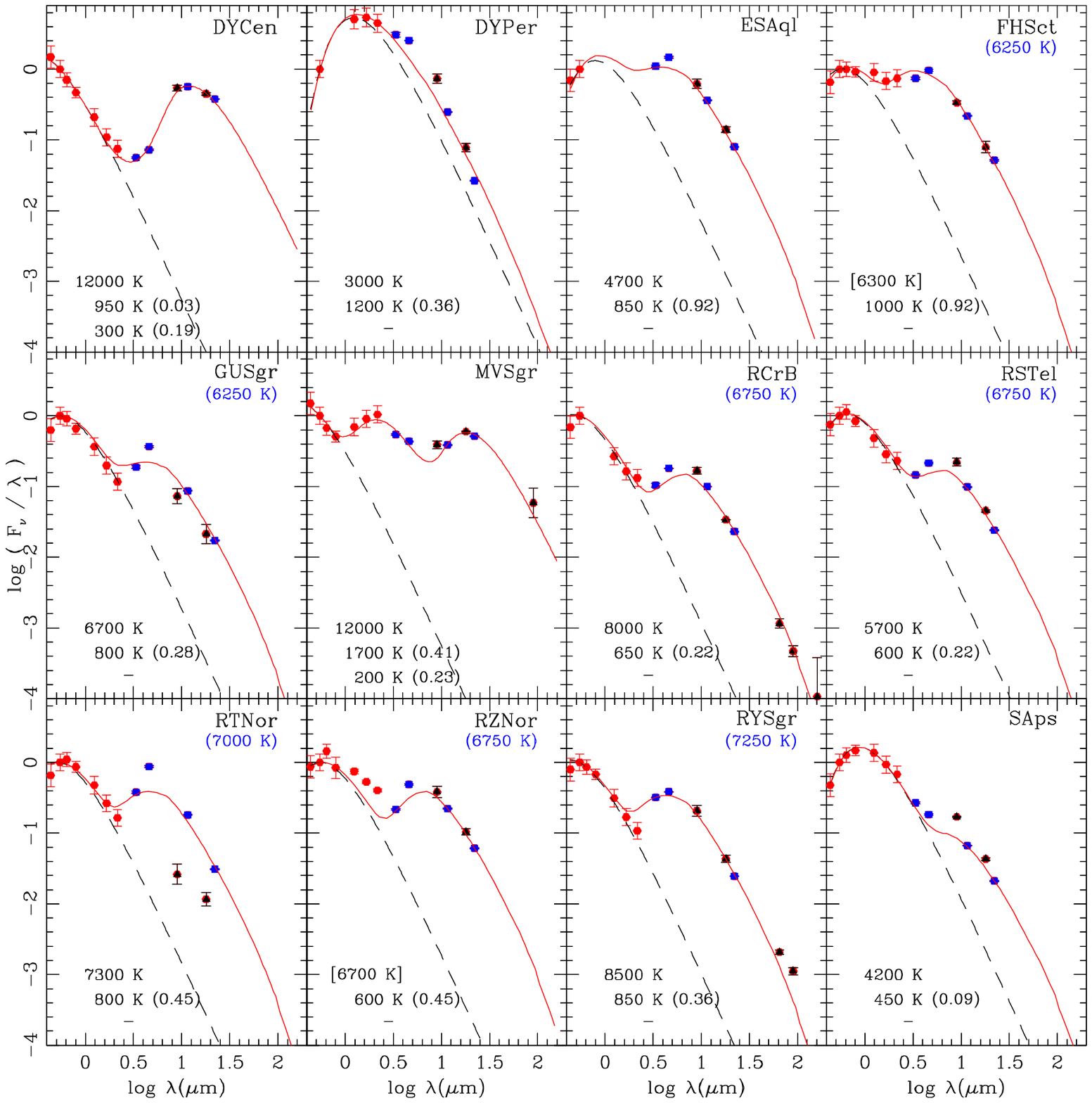}
\caption{Spectral energy distributions of known bright Galactic RCB stars, normalised to flux in V. Red dots represent fluxes in the optical (B, V, R and I) and the near-infrared (J, H and K) obtained from AAVSO and the ASAS, DENIS and 2MASS surveys. The blue and black dots represent mid-infrared fluxes from the WISE and AKARI surveys respectively. The red line is the best fit found with DUSTY models (see text for more details); the related effective temperatures found ($\sim10$\% level accuracy) are listed at the bottom left corner in the following order, from top to bottom : photosphere, first and second circumstellar shell. If the photosphere effective temperature is between square brackets, its value was fixed during the fit and corresponds to the photosphere effective temperature determined by \citet{2000A&A...353..287A}, indicated below the name for 17 RCB stars. The values in brackets on the right side of the shell temperatures correspond to the visual optical depth found. The broken black line represents a simple blackbody function with the photosphere temperature.}
\label{SED1}
\end{figure*}

\begin{figure*}
\centering
\includegraphics[scale=0.9]{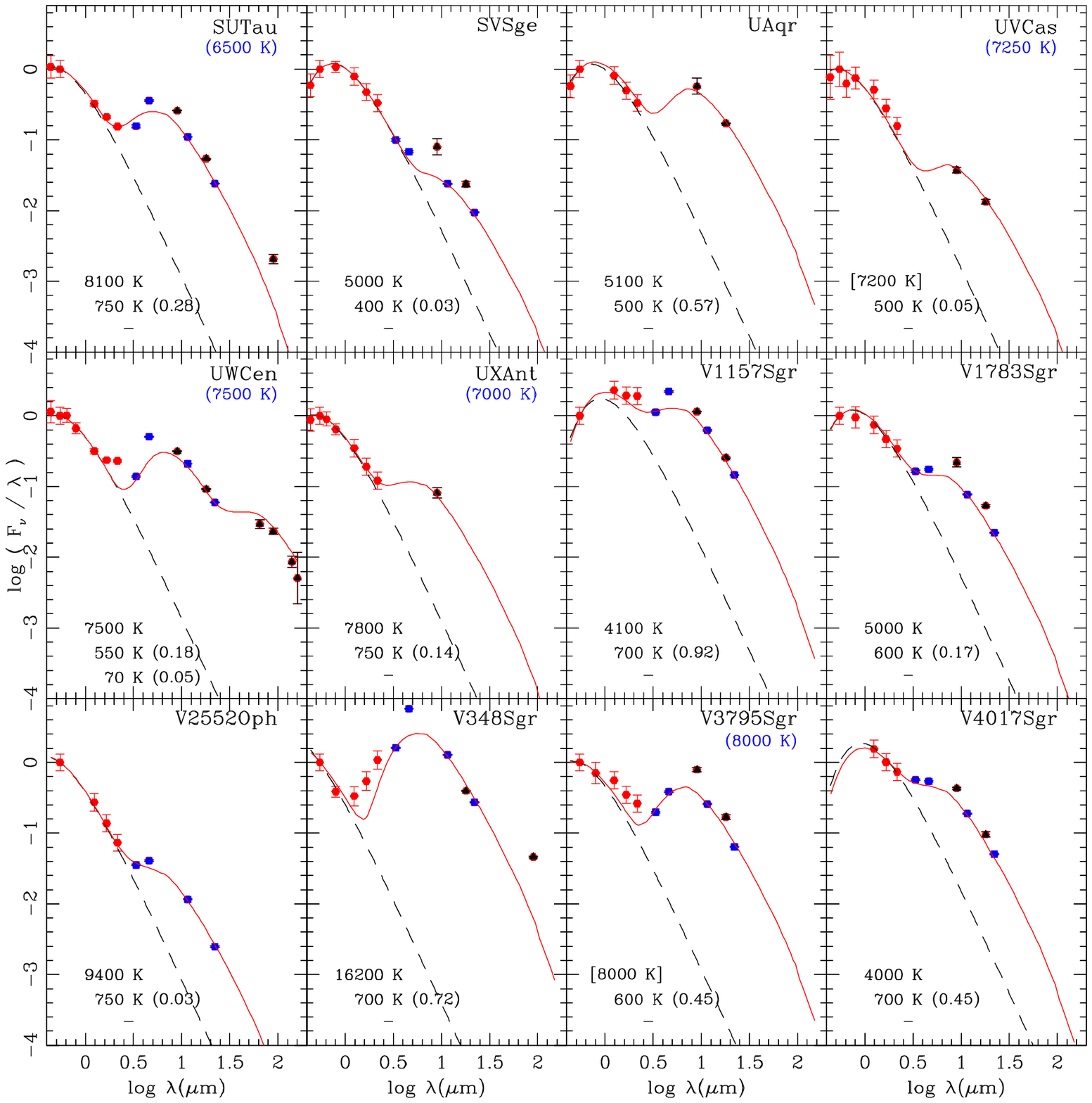}
\caption{Spectral energy distributions of known bright Galactic RCB stars, normalised to flux in V. Same caption as Figure~\ref{SED1}.}
\label{SED2}
\end{figure*}

\begin{figure*}
\centering
\includegraphics[scale=0.9]{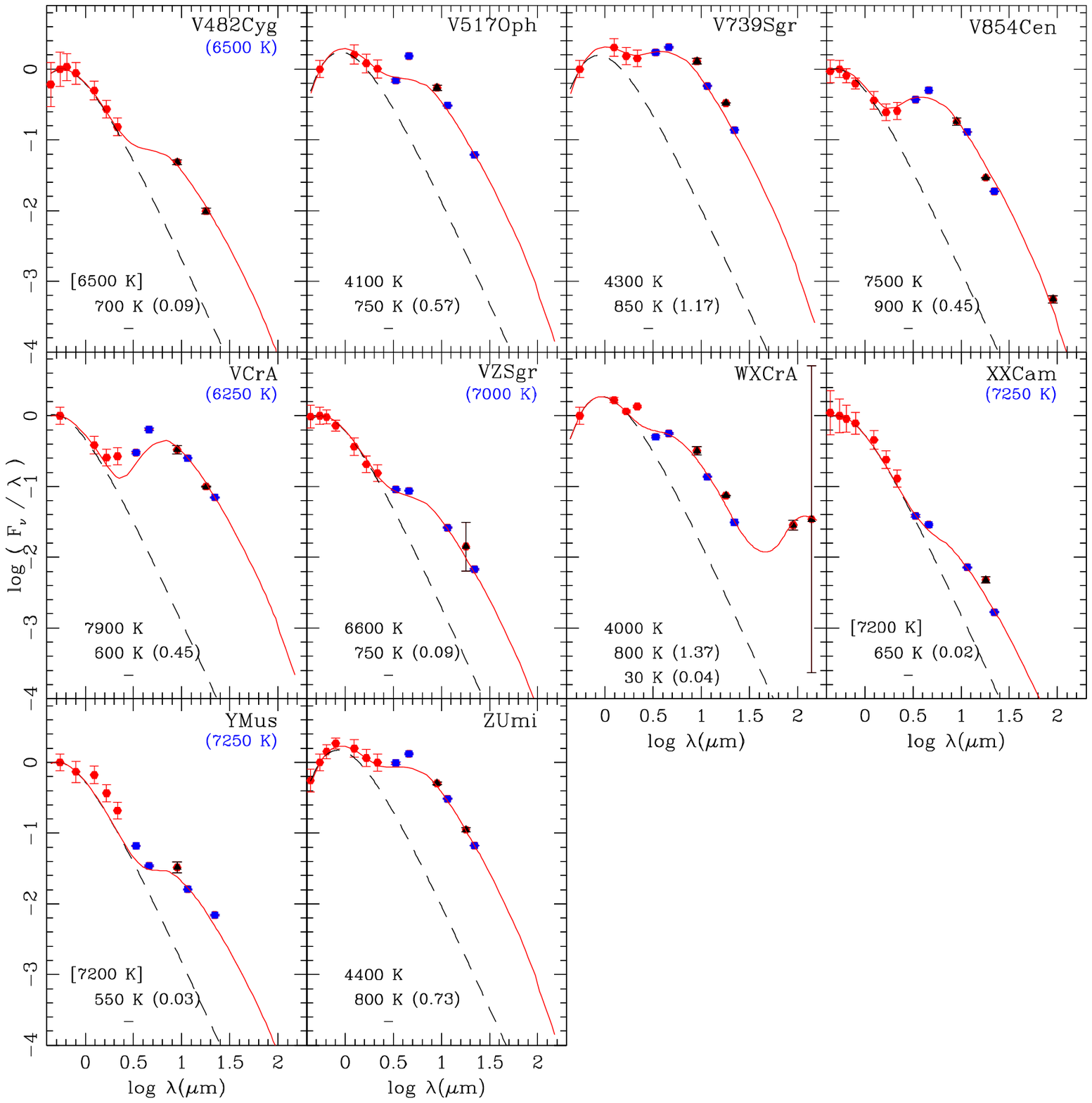}
\caption{Spectral energy distributions of known bright Galactic RCB stars, normalised to flux in V. Same caption as Figure~\ref{SED1}.}
\label{SED3}
\end{figure*}

\begin{figure*}
\centering
\includegraphics[scale=0.8]{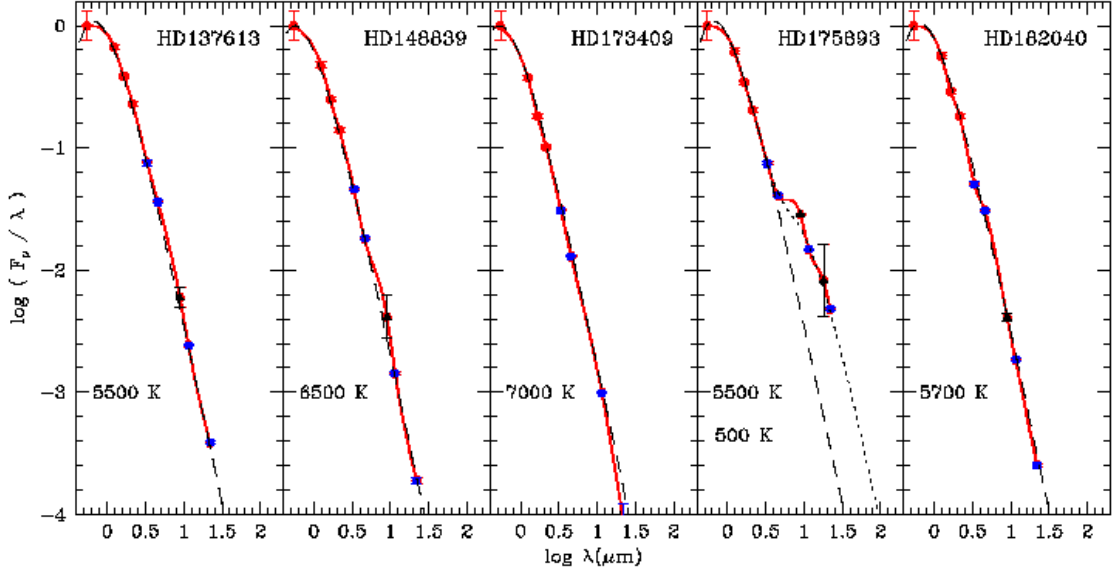}
\caption{Spectral energy distributions of the 5 HdC stars known, normalised to flux in V. Red dots represent fluxes in the optical and the near-infrared obtained from AAVSO and the ASAS, DENIS and 2MASS surveys. The blue and black dots represent the mid-infrared fluxes from the WISE and AKARI surveys respectively. The red line is simply a spline function that connects the different fluxes. The broken black line represents a simple blackbody function with the temperature of the photosphere. In the case of HD175893, the dotted line represents the sum of two blackbodies, the photosphere and the shell. The black bodies' effective temperatures of the photosphere (top) and the circumstellar shell (bottom) are listed at the bottom left corner.}
\label{SED4}
\end{figure*}

\begin{table*}
\caption{WISE magnitudes of catalogued RCB, HdC and DY Per stars.
\label{tab.WISE}}
\medskip
\centering
\begin{tabular}{lcccccccc}
\hline
\hline
 Name & [3.0] & $\sigma_{[3.0]}$ & [4.6] & $\sigma_{[4.6]}$ & [12] & $\sigma_{[12]}$ & [22] & $\sigma_{[22]}$  \\
\hline
\hline
& & & Galactic RCB stars & & & & &\\
\hline
\object{DY Cen} & 10.438 & 0.025 & 9.176 & 0.021 & 4.106 & 0.014 & 2.398 & 0.013 \\
\object{ES Aql} & 5.954 & 0.044 & 4.660$^b$ & 0.035 & 3.344 & 0.020 & 2.837 & 0.017 \\
\object{FH Sct} & 6.076 & 0.044 & 4.811$^b$ & 0.031 & 3.589 & 0.019 & 3.004 & 0.021 \\
\object{GU Sgr} & 5.993 & 0.051 & 4.289$^b$ & 0.039 & 3.021 & 0.019 & 2.624 & 0.016 \\
\object{MV Sgr} & 8.144 & 0.024 & 7.400 & 0.021 & 4.695 & 0.019 & 2.234 & 0.017 \\
\object{R CrB} & 3.505 & 0.067 & 1.920$^b$ & 0.002 & -0.273 & 0.003 & -0.833 & 0.019 \\
\object{RS Tel} & 7.075 & 0.031 & 5.672 & 0.027 & 3.676 & 0.018 & 3.057 & 0.021 \\
\object{RT Nor} & 5.877 & 0.044 & 3.987$^b$ & 0.035 & 2.856 & 0.020 & 2.628 & 0.018 \\
\object{RZ Nor} & 6.687 & 0.030 & 4.820$^b$ & 0.031 & 2.836 & 0.021 & 2.087 & 0.018 \\
\object{RY Sgr} & 2.615 & 0.004 & 1.434$^b$ & 0.001 & -- & -- & -0.564 & 0.012 \\
\object{S Aps} & 6.064 & 0.044 & 5.498 & 0.027 & 3.757 & 0.018 & 2.854 & 0.018 \\
\object{SU Tau} & 4.928 & 0.055 & 3.041$^b$ & 0.028 & 1.484 & 0.008 & 0.987 & 0.013 \\
\object{SV Sge} & 5.600 & 0.048 & 5.025 & 0.033 & 3.322 & 0.019 & 2.191 & 0.015 \\
\object{UW Cen} & 5.705 & 0.029 & 3.322$^b$ & 0.016 & 1.442 & 0.006 & 0.655 & 0.012 \\
\object{V1157 Sgr} & 6.291 & 0.038 & 4.582$^b$ & 0.031 & 3.118 & 0.022 & 2.549 & 0.020 \\
\object{V1783 Sgr} & 6.128 & 0.041 & 5.082 & 0.027 & 3.133 & 0.021 & 2.333 & 0.018 \\
\object{V2552 Oph} & 7.360 & 0.027 & 6.224 & 0.023 & 4.760 & 0.019 & 4.285 & 0.028 \\
\object{V348 Sgr} & 5.406 & 0.020 & 3.042$^b$ & 0.009 & 1.838 & 0.013 & 1.361 & 0.015 \\
\object{V3795 Sgr} & 7.153 & 0.029 & 5.439 & 0.025 & 3.045 & 0.020 & 2.410 & 0.017 \\
\object{V4017 Sgr} & 7.276 & 0.028 & 6.345 & 0.023 & 4.653 & 0.018 & 3.941 & 0.021 \\
\object{V517 Oph} & 4.969 & 0.058 & 3.121$^b$ & 0.033 & 2.031 & 0.018 & 1.624 & 0.016 \\
\object{V739 Sgr} & 6.448 & 0.037 & 5.283 & 0.028 & 3.818 & 0.019 & 3.225 & 0.021 \\
\object{V854 Cen} & 3.120 & 0.011 & 1.813$^b$ & 0.002 & 0.448 & 0.003 & 0.398 & 0.011 \\
\object{V CrA} & 6.008 & 0.050 & 4.211$^b$ & 0.046 & 2.387 & 0.012 & 1.624 & 0.013 \\
\object{VZ Sgr} & 7.405 & 0.028 & 6.474 & 0.019 & 4.938 & 0.019 & 4.260 & 0.027 \\
\object{WX CrA} & 6.206 & 0.041 & 5.100 & 0.031 & 3.798 & 0.017 & 3.259 & 0.023 \\
\object{XX Cam} & 5.431 & 0.053 & 4.760$^b$ & 0.033 & 3.427 & 0.019 & 2.867 & 0.018 \\
\object{Y Mus} & 8.133 & 0.025 & 7.855 & 0.022 & 5.846 & 0.018 & 4.612 & 0.025 \\
\object{Z Umi} & 5.982 & 0.045 & 4.673$^b$ & 0.030 & 3.429 & 0.018 & 2.933 & 0.017 \\
\object{MACHO-135.27132.51} & 8.337 & 0.024 & 7.033 & 0.019 & 5.043 & 0.013 & 4.323 & 0.031 \\
\object{MACHO-301.45783.9} & 8.540 & 0.027 & 7.177 & 0.021 & 5.351 & 0.018 & 4.578 & 0.030 \\
\object{MACHO-308.38099.66} & 7.647 & 0.023 & 6.495 & 0.021 & 4.829 & 0.016 & 4.071 & 0.025 \\
\object{MACHO-401.48170.2237} & 5.901 & 0.044 & 4.576$^b$ & 0.035 & 3.407 & 0.019 & 2.858 & 0.028 \\
\object{EROS2-CG-RCB-1} & 6.666 & 0.039 & 5.036 & 0.026 & 3.216 & 0.019 & 2.334 & 0.018 \\
\object{EROS2-CG-RCB-3} & 5.768 & 0.047 & 3.954$^b$ & 0.068 & 3.099 & 0.018 & 2.585 & 0.020 \\
\object{EROS2-CG-RCB-4} & 6.411 & 0.036 & 5.080 & 0.030 & 3.635 & 0.017 & 2.998 & 0.027 \\
\object{EROS2-CG-RCB-5} & 6.319 & 0.035 & 4.867$^b$ & 0.030 & 3.784 & 0.015 & 3.144 & 0.022 \\
\object{EROS2-CG-RCB-6} & 7.140 & 0.031 & 5.933 & 0.025 & 4.314 & 0.020 & 3.670 & 0.034 \\
\object{EROS2-CG-RCB-7} & 7.210 & 0.031 & 6.143 & 0.023 & 4.628 & 0.018 & 3.985 & 0.029 \\
\object{EROS2-CG-RCB-8} & 7.050 & 0.032 & 5.795 & 0.026 & 4.189 & 0.019 & 3.595 & 0.036 \\
\object{EROS2-CG-RCB-9} & 7.250 & 0.028 & 5.406 & 0.027 & 3.423 & 0.018 & 2.649 & 0.023 \\
\object{EROS2-CG-RCB-10} & 6.739 & 0.036 & 4.485$^b$ & 0.039 & 2.546 & 0.023 & 1.958 & 0.022 \\
\object{EROS2-CG-RCB-11} & 6.581 & 0.035 & 5.456 & 0.026 & 4.138 & 0.016 & 3.523 & 0.024 \\
\object{EROS2-CG-RCB-12} & 8.067 & 0.026 & 7.380 & 0.022 & 6.452 & 0.019 & 6.263 & 0.055 \\
\object{EROS2-CG-RCB-13} & 6.839 & 0.034 & 5.541 & 0.026 & 3.985 & 0.016 & 3.378 & 0.026 \\
\object{EROS2-CG-RCB-14} & 6.375 & 0.034 & 4.695$^b$ & 0.031 & 3.292 & 0.019 & 2.678 & 0.019 \\
\hline
& & & Magellanic RCB stars & & & & &\\
\hline
\object{MACHO-12.10803.56} & 10.287 & 0.025 & 9.367 & 0.021 & 7.744 & 0.020 & 7.208 & 0.055 \\
\object{EROS2-LMC-RCB-4} & 10.001 & 0.025 & 8.551 & 0.021 & 6.637 & 0.018 & 6.139 & 0.034 \\
\object{EROS2-LMC-RCB-5} & 11.854 & 0.025 & 11.365 & 0.022 & 8.623 & 0.022 & 7.437 & 0.067 \\
\object{EROS2-LMC-RCB-6} & 10.604 & 0.025 & 9.082 & 0.022 & 6.857 & 0.018 & 6.102 & 0.035 \\
\hline
& & & & & & & &\\
& & & HdC stars & & & & &\\
\hline
\object{HD137613} & 5.090 & 0.059 & 4.909$^b$ & 0.033 & 5.000 & 0.018 & 4.854 & 0.028 \\
\object{HD148839} & 6.787 & 0.033 & 6.804 & 0.023 & 6.731 & 0.019 & 6.770 & 0.063 \\
\object{HD173409} & 8.095 & 0.018 & 8.067 & 0.019 & 8.021 & 0.025 & 8.508 & 0.349 \\
\object{HD175893} & 7.165 & 0.029 & 6.832 & 0.021 & 5.104 & 0.019 & 4.175 & 0.025 \\
\object{HD182040} & 4.953 & 0.054 & 4.511$^b$ & 0.033 & 4.730 & 0.019 & 4.739 & 0.032 \\
\hline
& & & DY Per stars & & & & &\\
\hline
\object{DY Per} & 3.030 & 0.089 & 2.254$^b$ & 0.007 & 1.941 & 0.009 & 2.221 & 0.022 \\
\object{EROS2-CG-RCB-2} & 8.189 & 0.035 & 7.856 & 0.030 & 7.388 & 0.045 & 7.376 & null \\
\object{MACHO-15.10675.10} & 9.525 & 0.023 & 9.344 & 0.020 & 8.751 & 0.021 & 8.679 & 0.135 \\
\object{EROS2-LMC-DYPer-5} & 10.304 & 0.024 & 10.178 & 0.021 & 9.616 & 0.027 & 9.091 & 0.240 \\
\hline
\hline
\multicolumn{9}{l}{$^b$[4.6] magnitude affected by bias (see Fig.~\ref{4.6ObsModel}) } \\
\end{tabular}
\end{table*}

\begin{table*}
\caption{AKARI fluxes (mJy) of catalogued bright Galactic RCB stars and DY Persei.
\label{tab.AKARI}}
\medskip
\centering
\begin{tabular}{lcccccccccccc}
\hline
\hline
 Name & [9.0] & $\sigma_{[9.0]}$ & [18] & $\sigma_{[18]}$ & [65] & $\sigma_{[65]}$ & [90] & $\sigma_{[90]}$ & [140] & $\sigma_{[140]}$ & [160] & $\sigma_{[160]}$  \\
\hline
\hline
\object{DY Cen} & 0.538 & 0.022 & 0.889 & 0.016 & -- & -- & -- & -- & -- & -- & -- & -- \\
\object{ES Aql} & 1.935 & 0.127 & 0.875 & 0.036 & -- & -- & -- & -- & -- & -- & -- & -- \\
\object{FH Sct} & 1.397 & 0.032 & 0.661 & 0.053 & -- & -- & -- & -- & -- & -- & -- & -- \\
\object{GU Sgr} & 1.290 & 0.140 & 0.751 & 0.100 & -- & -- & -- & -- & -- & -- & -- & -- \\
\object{MV Sgr} & 0.330 & 0.019 & 1.008 & 0.008 & -- & -- & 0.496 & 0.103 & -- & -- & -- & -- \\
\object{R CrB} & 52.990 & 2.440 & 21.480 & 0.029 & 2.656 & 0.170 & 1.494 & 0.114 & -- & -- & 0.606 & 0.333 \\
\object{RS Tel} & 1.870 & 0.102 & 0.767 & 0.011 & -- & -- & -- & -- & -- & -- & -- & -- \\
\object{RT Nor} & 0.257 & 0.037 & 0.228 & 0.021 & -- & -- & -- & -- & -- & -- & -- & -- \\
\object{RZ Nor} & 3.119 & 0.252 & 1.695 & 0.076 & -- & -- & -- & -- & -- & -- & -- & -- \\
\object{RY Sgr} & 48.000 & 3.660 & 20.180 & 1.020 & 3.495 & 0.104 & 2.605 & 0.139 & -- & -- & -- & -- \\
\object{S Aps} & 1.971 & 0.016 & 1.001 & 0.021 & -- & -- & -- & -- & -- & -- & -- & -- \\
\object{SU Tau} & 14.720 & 0.050 & 6.161 & 0.046 & 0.351 & -- & 1.179 & 0.080 & -- & -- & -- & -- \\
\object{SV Sge} & 3.849 & 0.441 & 2.293 & 0.095 & -- & -- & -- & -- & -- & -- & -- & -- \\
\object{U Aqr} & 1.260 & 0.142 & 0.750 & 0.002 & -- & -- & -- & -- & -- & -- & -- & -- \\
\object{UV Cas} & 1.125 & 0.046 & 0.802 & 0.026 & -- & -- & -- & -- & -- & -- & -- & -- \\
\object{UW Cen} & 9.760 & 0.070 & 5.697 & 0.047 & 6.645 & 0.413 & 7.302 & 0.332 & 4.179 & 0.349 & 2.809 & 1.020 \\
\object{UX Ant} & 0.101 & 0.007 & -- & -- & -- & -- & -- & -- & -- & -- & -- & -- \\
\object{V1157 Sgr} & 2.569 & 0.004 & 1.145 & 0.015 & -- & -- & -- & -- & -- & -- & -- & -- \\
\object{V1783 Sgr} & 3.918 & 0.252 & 1.900 & 0.032 & -- & -- & -- & -- & -- & -- & -- & -- \\
\object{V348 Sgr} & -- & -- & 2.813 & 0.028 & -- & -- & 1.638 & 0.018 & -- & -- & -- & -- \\
\object{V3795 Sgr} & 4.574 & 0.144 & 1.951 & 0.069 & -- & -- & -- & -- & -- & -- & -- & -- \\
\object{V482 Cyg} & 1.128 & 0.037 & 0.459 & 0.020 & -- & -- & -- & -- & -- & -- & -- & -- \\
\object{V517 Oph} & 6.756 & 0.243 & -- & -- & -- & -- & -- & -- & -- & -- & -- & -- \\
\object{V739 Sgr} & 1.638 & 0.065 & 0.840 & 0.001 & -- & -- & -- & -- & -- & -- & -- & -- \\
\object{V854 Cen} & 22.970 & 1.170 & 7.364 & 0.033 & -- & -- & 0.705 & 0.036 & -- & -- & -- & -- \\
\object{V CrA} & 3.606 & 0.210 & 2.166 & 0.013 & -- & -- & -- & -- & -- & -- & -- & -- \\
\object{VZ Sgr} & -- & -- & 0.283 & 0.098 & -- & -- & -- & -- & -- & -- & -- & -- \\
\object{WX CrA} & 1.742 & 0.097 & 0.810 & 0.001 & -- & -- & 1.548 & 0.105 & 2.919 & 6.340 & -- & -- \\
\object{XX Cam} & -- & -- & 1.391 & 0.065 & -- & -- & -- & -- & -- & -- & -- & -- \\
\object{Y Mus} & 0.230 & 0.017 & -- & -- & -- & -- & -- & -- & -- & -- & -- & -- \\
\object{Z Umi} & 1.724 & 0.029 & 0.765 & 0.024 & -- & -- & -- & -- & -- & -- & -- & -- \\
\hline
\object{DY Per} & 12.210 & 0.823 & 2.600 & 0.157 & -- & -- & -- & -- & -- & -- & -- & -- \\
\hline
\hline
\end{tabular}
\end{table*}

\begin{sidewaystable*}
\caption{First 10 rows of the published catalogue.
\label{tab.short_version}}
\centering
\begin{tabular}{lccccccccccccccc}
\hline
ID  & Ra & Dec & l & b & \multicolumn{2}{c}{WISE} & \multicolumn{2}{c}{WISE} & \multicolumn{2}{c}{WISE} & \multicolumn{2}{c}{WISE} & \multicolumn{2}{c}{2MASS} \\
 & (deg) & (deg) & (deg) & (deg) & [3.0] & $\sigma_{[3.0]}$ & [4.6] & $\sigma_{[4.6]}$ & [12] & $\sigma_{[12]}$ & [22] & $\sigma_{[22]}$ & J mag & $\sigma_{J}$ \\
\hline
\hline
1 &	105.4079971 & -41.4713669 & 251.85892 & -15.80245 & 4.852 & 0.058 & 3.017 & 0.016 & 1.736 & 0.017 & 1.197 & 0.016 & 11.324 & 0.024 \\
2 &	259.1376648 & -33.8375053 & 352.49854 & 2.49227 & 8.529 & 0.029 & 6.471 & 0.022 & 3.917 & 0.016 & 2.809 & 0.019 & 17.348 & -99 \\
3 &	109.6635208 & -24.9482384 & 238.16979 & -5.55101 & 9.701 & 0.026 & 8.951 & 0.021 & 7.202 & 0.020 & 6.375 & 0.085 & 11.004 & 0.022 \\
4 &	107.7039337 & -6.3668070 & 220.80645 & 1.38275 & 6.906 & 0.029 & 4.715 & 0.034 & 2.462 & 0.022 & 1.557 & 0.021 & 13.736 & 0.030 \\
5 &	273.2704468 & -25.0299644 & 6.43576 & -3.35900 & 10.990 & 0.078 & 11.098 & 0.085 & 9.010 & 0.158 & 8.046 & 0.329 & 12.274 & -99 \\
6 &	258.5602112 & -21.4371471 & 2.42136 & 10.03461 & 6.800 & 0.032 & 5.829 & 0.022 & 4.418 & 0.019 & 4.017 & 0.023 & 9.783 & 0.024 \\
7 &	273.8857422 & -25.5371208 & 6.25302 & -4.08940 & 10.657 & 0.050 & 10.018 & 0.040 & 8.367 & 0.031 & 7.509 & 0.178 & 12.799 & -99 \\
8 &	273.0597229 & -25.4018135 & 6.01729 & -3.36891 & 10.507 & 0.068 & 9.501 & 0.038 & 7.216 & 0.025 & 6.009 & 0.070 & 12.616 & 0.029 \\
9 &	272.6348572 & -25.3729763 & 5.85848 & -3.01816 & 9.329 & 0.035 & 9.041 & 0.026 & 8.114 & 0.030 & 8.043 & 0.289 & 10.410 & 0.021 \\
10 &	273.3207092 & -25.5264320 & 6.02011 & -3.63545 & 9.215 & 0.028 & 7.982 & 0.021 & 6.220 & 0.018 & 5.708 & 0.043 & 14.477 & 0.068 \\
\hline
\multicolumn{16}{c}{}\\
\multicolumn{16}{c}{ Colunms continued }\\
\hline
 \multicolumn{2}{c}{2MASS} & \multicolumn{2}{c}{2MASS} & \multicolumn{2}{c}{DENIS} & \multicolumn{2}{c}{DENIS} & \multicolumn{2}{c}{DENIS} & \multicolumn{5}{c}{USNO-B1} & SIMBAD \\
 H mag & $\sigma_{H}$ & K mag & $\sigma_{K}$ & I mag & $\sigma_{I}$ & J mag & $\sigma_{J}$ & K mag & $\sigma_{K}$ & B1 mag & B2 mag & R1 mag & R2 mag & I mag & Class. \\
\hline
\hline
 9.154 & 0.024 & 7.287 & 0.027 &	 15.768 & 0.06 & 12.201 & 0.06 & 8.076 & 0.06 & -99 & 17.06 & -99 & 16.97 & 15.91 &	C* \\
 16.261 & -99 & 13.424 & 0.050 &	 -99 & -99 & -99 & -99 & 12.049 & 0.11 & -99 & -99 & -99 & -99 & -99 &	\_ \\
 10.798 & 0.026 & 10.549 & 0.023 &	 -99 & -99 & -99 & -99  &-99 & -99 & -99 & -99 & -99 & -99  &-99 &	*iC \\
 10.807 & 0.027 & 8.466 & 0.023 &	 -99 & -99 & 13.650 & 0.09 & 8.349 & 0.07 & -99 & -99 & -99 & -99 & -99 &	\_ \\
 11.694 & 0.049 & 11.312 & 0.038 &	 -99 & -99 & -99 & -99 & 11.294 & 0.08 & -99 & -99 & -99 & -99 & -99 &	\_ \\
 9.036 & 0.024 & 8.479 & 0.021 &	 10.806 & 0.04 & 9.393 & 0.07 & 8.261 & 0.25 & 14.13 & 11.90 & 15.26 & 11.89 & 10.42 &	C* \\
 12.844 & 0.064 & 12.196 & 0.034 &	 -99 & -99 & -99 & -99 & -99 & -99 & -99 & -99 & -99 & -99 & -99 &	\_ \\
 12.018 & 0.027 & 11.493 & 0.021 &	 14.293 & 0.04 & 12.961 & 0.06 & 11.776 & 0.09 & 16.67 & 14.12 & 16.13 & 15.39 & 13.81	& \_ \\
 10.008 & 0.025 & 9.735 & 0.021 &	 12.816 & 0.32 & 11.404 & 0.40 & 10.819 & 0.55 & 14.20 & 12.67 & 14.15 & 13.05 & 13.38	& \_ \\
 12.586 & 0.094 & 11.067 & 0.039 &	 16.327 & 0.09 & 13.232 & 0.13 & 10.515 & 0.13 & -99 & -99 & -99 & -99 & -99 &	V* \\
\hline
\end{tabular}
\end{sidewaystable*}

\begin{acknowledgements}

I thank personally Tony Martin-Jones, Julia Jane Karrer and Mike Bessell for their careful reading and comments.

This publication makes use of data products from the Wide-field Infrared Survey Explorer, which is a joint project of the University of California, Los Angeles, and the Jet Propulsion Laboratory/California Institute of Technology, funded by the National Aeronautics and Space Administration.
This publication also makes use of data products from the Two Micron All Sky Survey, which is a joint project of the University of Massachusetts and the Infrared Processing and Analysis Centre, California Institute of Technology, funded by the National Aeronautics and Space Administration and the National Science Foundation. The DENIS data have also been used. DENIS is the result of a joint effort involving human and financial contributions of several Institutes mostly located in Europe. It has been supported financially mainly by the French Institut National des Sciences de l'Univers, CNRS, and French Education Ministry, the European Southern Observatory, the State of Baden-Wuerttemberg, and the European Commission under networks of the SCIENCE and Human Capital and Mobility programs, the Landessternwarte, Heidelberg and Institut d'Astrophysique de Paris.
This article is also based on observations with AKARI, a JAXA project with the participation of ESA. 

Finally, I thank the AAVSO association which provides variable star light curves with observations contributed by observers worldwide, and to the ASAS south survey that provides light curves of bright stars from the entire southern sky.

\end{acknowledgements}

\bibliographystyle{aa}
\bibliography{WISE_RCB}


\end{document}